# Is climate variability the result of frequency modulation by the solar cycle? Evidence from the El Nino Southern Oscillation, Australian climate, Central England Temperature, and reconstructed solar activity and climate records.

Ian R. Edmonds   iredmonds@aapt.net.au

**Highlights:** Solar cycle length, frequency modulation; modelling El Nino Southern Oscillation, Central England Temperature, Paleo-climate reconstructions

**Abstract**   Oceanic atmospheric oscillations and climate variability are tightly linked and both exhibit broad band spectral content that ranges, with roughly equal strength, from bi-annual to centennial periodicity.  The explanation for variability based on the integration of weather noise leads to a spectral content heavily weighted to low frequencies; explaining the variability as resulting from solar forcing leads to a narrow band, approximately eleven year period, spectral content. In both cases the spectral content is incompatible with the observed spectrum. It is known that the El Nino Southern Oscillation is frequency modulated, i.e. the time interval between successive El Nino's varies on an approximately centenary scale. In this paper we develop a model of the El Nino Southern Oscillation responding to the slowly changing frequency of the solar cycle. This results in a frequency modulated oscillation, the spectrum of which is intrinsically broad and flat and therefore compatible with the observed spectrum. Fortunately, the change in frequency of the solar cycle with time has been reconstructed from tree ring data for the last millennium.  It is possible to identify time intervals when the frequency was dominated by a single frequency in which case the model oscillation is relatively simple. The decadal component of the model time variation was shown to correlate closely with the decadal components of observed El Nino Southern Oscillation and climate variability.  A characteristic of a frequency modulated variable, the equal spacing of spectral peaks, was utilised via a double Fourier transform method to recover solar cycle periodicity from instrumental and reconstructed climate records, with the recovered periodicity and the known periodicity of the solar cycle in good agreement. The concept outlined provides a new way of viewing and assessing the Sun – climate connection.

## 1 Introduction

Climate change is conventionally measured in terms of variation in temperature and rainfall. Explanation of the variation is challenging as the variability of climate is complex, having a broad spectral content that ranges from bi-annual to millennial. One concept, controversial, but of very long standing, is that climate variability on Earth is connected to cyclic variability of the Sun. Solar variability is associated with the approximately 22 year cycle during which the magnetic field of the Sun reverses in direction. This cycle, the Hale cycle, leads to the recurrent formation and waning of sunspots in an approximately 11 year cycle, the Schwabe cycle, known more generally as the solar activity cycle or solar cycle. This cyclic recurrence of sunspots leads to other forms of cyclic variability in variables such as solar irradiance, solar wind, interplanetary magnetic field and cosmic ray flux. The observed sunspot cycle, has been extensively studied by solar physicists, for example (Frolich and Lean 1998, Feynman and Ruzamaikin 2014), proxy versions of the sunspot cycle reconstructed from tree rings and ice cores extend over millennia and it is known that the sunspot cycle varies widely in amplitude between grand maxima and grand minima and moderately in frequency, i.e. the period of the solar cycle is confined to a narrow band between 8 and 13 years, Usoskin et al (2021). However, why the solar cycle varies as observed is still not well understood, Charbonneau (2020). Climate scientists have sought an understanding of how solar variability modulates climate variability, (Haigh 1996, Rind 2002, Labitske and Mathes 2003, Gray et al 2010, Bal et al 2011, Scafetta and Bianchini 2023). However, climate variability is much more complex than solar variability. If there was a direct relationship between the solar cycle and climate variability, i.e. if climate was modulated by the amplitude of the solar cycle, climate variability would be as narrow band as the solar cycle. There are two basic types of modulation of an oscillatory system, amplitude modulation where the amplitude of the oscillator is proportional to amplitude of the modulator, and frequency modulation where the frequency of the oscillator is proportional to the amplitude of the modulator. In contrast to amplitude modulation a frequency modulated signal is broad band and complex. A solar/climate connection should follow the sequence, solar variability – atmospheric/oceanic oscillation variability – climate variability. This article considers the possibility that climate variability results from the frequency modulation of climate oscillatory systems such as the El Nino Southern Oscillation (ENSO) by a solar cycle that varies in frequency. The development of this paper proceeded from an interest in why the decadal component of Australian rainfall had nearly tripled in amplitude since 1960. This led to an attempt at understanding, in a more general way, how climate variability was influenced by solar variability. Thus the first sections of the paper assess decadal variability in Australian climate and the connection with decadal variation in ENSO, while the later sections consider the frequency variation of the solar cycle and the resultant frequency modulation of ENSO and climate variables.

This work was facilitated by the use of a method of fast band pass filtering, the inverted notch filter, (INF), as described briefly in Methods, Section 2. Section 3 provides evidence, using online Bureau of Meteorolgy, (BOM), data, that decadal components in Australian climate variables are, currently, strong and increasing. Section 4 demonstrates the high correlation of Australian climate variability to ENSO. Section 5 describes the frequency variation of the solar cycle, Section 6 formulates the frequency modulation of a simple harmonic oscillator model of ENSO, and compares the time variations of the model and ENSO in the decadal range. Section 7 provides supporting evidence for the frequency modulated ENSO model, Section 8 is a discussion and a conclusion.

**2. Methods** Fast Fourier Transforms (FFT) were performed with the DPLot application. A fast method of band pass filtering, called here inverted notch filtering (INF), was performed by applying the Press notch filter, Press et al (2007), that is included in the DPlot application and subtracting the notch filtered data from the original data to achieve a band pass filtered version of the data. Throughout this paper the bandwidth of the notch filter was set to 10% of the centre frequency of the band.

### 3. The decadal cycle in Australian climate variables.

**3.1 Australian rainfall.** Dorothea Mackellar's description of Australia as "a land of drought and flooding rains" is as relevant today as it was in 1908 with the Millennium Drought and the 2011 floods amongst many recent examples of severe or extreme rainfall events. Within the literature, recent extremes of high temperature and high rainfall in Australia have been tentatively attributed to the effects of the approximately one degree centigrade increase in global temperature since 1850, for example, (Cai et al 2021, Cai et al 2023). However, extremes at both ends of the variable range have recently been experienced, with extremes of low temperature as well as low rainfall. It is debatable whether the fairly monotonic increase in global temperature since 1850 would lead to both positive and negative extremes in climate variables. Therefore it is useful to examine if other influences like solar activity could give rise to the positive and negative climate extremes of recent occurrence. Absence of this type of work by Australian climate scientists is notable with, apparently, only a few reports, (Wasko and Sharma 2009a, 2009b, Kidson 1925, Baker 2008, Bowen 1975), examining the effect of the solar cycle on rainfall. Wasko and Sharma demonstrated that when General Circulation Models included decadal solar forcing, the resulting global atmospheric moisture content exhibited a high power at the decadal period and, as a result, they concluded that decadal solar forcing could have a large effect on drought severity. The Australian Bureau of Meteorology (BOM) provides extensive time series data on its website that includes the main climate variables, rainfall, cloud cover and temperature. One objective of this paper is to show that Australian climate variables, in particular rainfall, do exhibit strong decadal periodicity. A second and more challenging objective of this paper was to assess whether the observed decadal periodicity is associated with the solar cycle. The decadal cycle in Australian rainfall was studied by Meinke et al (2005) with the objective of assessing whether the cycle is due to the solar activity or due to internal noise. For similar reasons Asten and McCracken (2022) assessed whether the Gleissberg ~ 90 year solar cycle was evident in Australian rainfall.

This section uses cloud, rainfall and temperature anomaly data downloaded from the Climate Time Series site the BOM provides online to the public at http://www.bom.gov.au/climate/change/. The site provides data for the six regions in Australia and for Australia as a whole. The rainfall data is available from 1900 to 2023, the cloud data from 1957 to 2014, and the temperature data from 1910 to 2023. Figure 1A shows the Australian rainfall anomaly, blue broken line, from 1900 to 2023. Also shown are the ~ 11 year period component, the ~ 32 year period component, and the linear trend. The components were obtained by the inverse notch filter (INF) method. The increasing amplitude of the ~ 11 year component with time is evident. In recent years the peak to peak range of the ~11 year component exceeds the extent of the change in the linear trend in Australian rainfall as measured over the entire 123 year record and is, therefore a very significant contributor to rainfall variation.

Figure 1B shows the periodogram of the rainfall anomaly obtained by the FFT after removing the linear trend from the anomaly. In this paper we are interested in the long term variation of Australian rainfall. There are four significant long term periodicities evident in the periodogram, labelled in Figure 1B as 8.5, 11.6, 18.3 and 32 years, the values labelled corresponding to the period at the highest point of each periodicity. However, there is also strong periodicity evident at about 5.5, 4.7, 3.5 and 2.7 year period. The broad and essentially flat spectrum is characteristic of the spectral content of climate variability generally. Evidently, the most prominent periodicity in rainfall corresponds to the approximately decadal periodicity labelled, in Figure 1B, at the 11.6 year period point. We will show below, in the section that discusses the connection between solar cycle length, ENSO, and Australian rainfall that the average period of the solar cycle as measured by the sunspot number, SSN, during the 1900 to 2023 interval is 10.7 years, a period somewhat shorter than the period of the cycle evident at 11.6 years in the spectrum of rainfall. The 10.7 year average period of the SSN is indicated by the dotted reference line in Figure 1B and the difference in periodicity of rainfall and the solar cycle would immediately preclude any direct connection between solar activity and this climate variable.

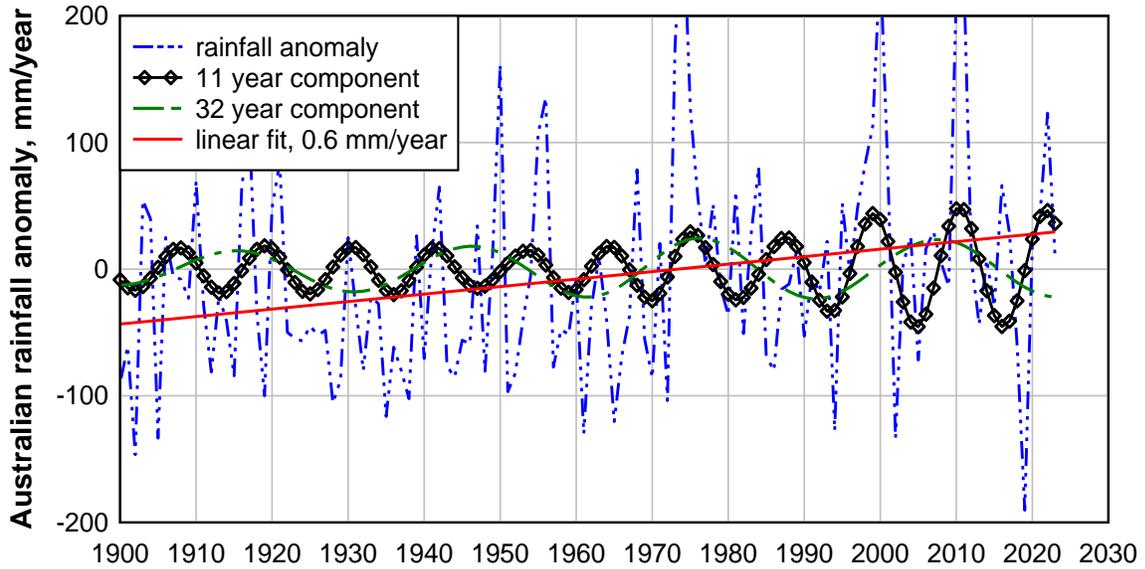

(A)

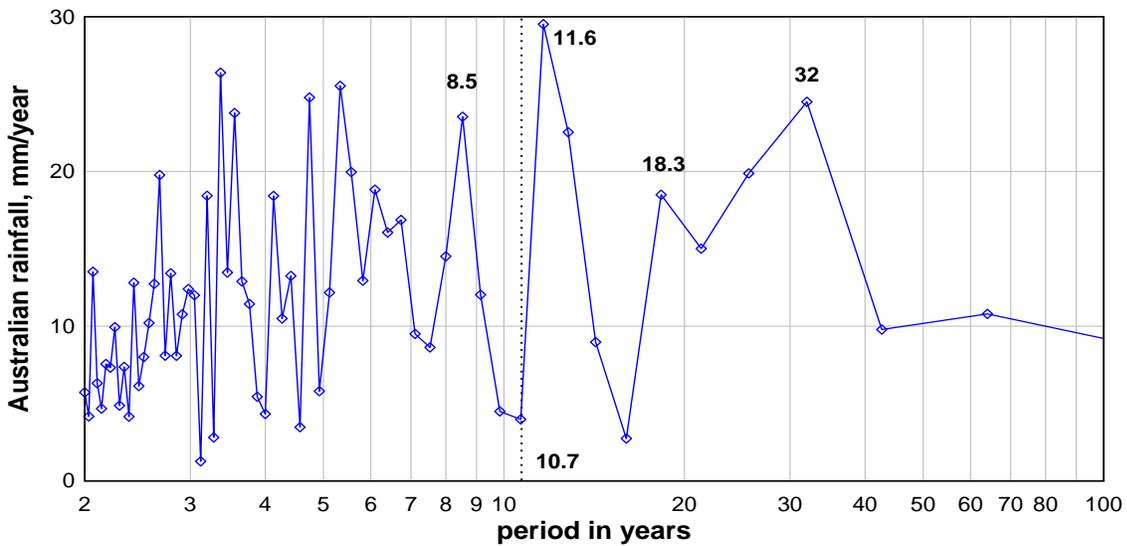

(B)

**Figure 1. (A).The Australian rainfall anomaly, blue broken line, and the linear fit of 0.6 mm/year, red line. The approximately 11 year period, diamond symbols, and the 32 year period components of rainfall are also shown. (B). Periodogram of Australian rainfall. The periodicity in years at the highest points of the four long period components is labelled. The average period, 10.7 years, of the solar cycle is indicated by the dotted reference line.**

**3.2 Regional analysis of Australian rainfall.** In presenting long term climate data the BOM divides Australia into six regions, Eastern, Northern, Southern, South Eastern, and South Western Australia, and the Murray Darling Basin, as well as Australia as a whole. Figure 2A presents the FFT spectrums of rainfall in each of the six regions. The rainfall periodicities that appear consistently in all regions are marked by reference lines and labelled by the corresponding period in years. The strongest consistent peak is here marked with a reference line at 12 year period. The other consistent periodicities are at 32, 20, 5.7, 4.7 and 3.5 years. Figure 2B shows the time variations of the ~11 year period component of rainfall for each of the six regions. It is noticeable that, apart from the component for the South Western Australia region, the ~11 year period components from each region are consistently strong and in-phase except for the time interval 1920 – 1960.

It is notable that, the amplitude of the ~ 11 year component decreases from 1900 to 1930, then from 1950 onwards, the amplitude of the component is strongly increasing with time in all regions apart from the South Western region. Consequently, the recent contribution of the ~11 year component to rainfall in most regions is large. For example, in the Eastern Australian region the average level of rainfall, 1900 to 2023, is 600 mm/year and, from Figure 2B, the recent cycles in the ~11 year period component of rainfall swing through about 200 mm/year, a swing of about 33% relative to the mean level. Clearly the ~11 year cycle in rainfall has a strong influence on whether Eastern Australia is in drought or flood.

It is evident from Figure 2B that the decadal period rainfall variation in the first half of the BOM records, 1900 to 1960 is much smaller than the variation in the second half, 1961 to 2023. The distinct difference in the spectral content of the two halves of the record is illustrated, later in the paper, in Figures 10A and 10B. There is clearly a shift from shorter periodicity, higher frequency rainfall events, in the first half of the record to longer periodicity, lower frequency rainfall events between in the second half of the record.

**3.3 The decadal cycle in Australian temperature and cloud cover.** The data plotted in Figures 2A and 2B indicate that observations from the Eastern region of Australia exhibit the strongest decadal component in rainfall. A direct relationship is expected between cloud cover and rainfall. It is expected that decadal components in cloud cover and temperature would be correlated, positively and negatively respectively, with decadal components in rainfall, Hope and Watterson (2018). Figure 3A illustrates that, for the Eastern Australian region, the expected correlations are nearly exact. An inverse relationship is expected between rainfall and daily maximum temperature due to the effect of cloud cover on solar irradiance and the cooling effect of rain. Figure 3A compares the ~ 11 year components Eastern Australian rainfall, daily maximum temperature, ($T_{MAX}$) and cloud cover as obtained from the BOM online site. There is an almost exact inverse relationship of the decadal components of rainfall and TMAX. Also, the amplitudes of the rainfall, cloud cover and the temperature components are increasing markedly from about 1950 onwards. To illustrate, the average $T_{MAX}$ 11 year component amplitude before 1950 is about 0.07 C; while the post 2000 the average amplitude is about 0.25 C, an increase by about three hundred percent.

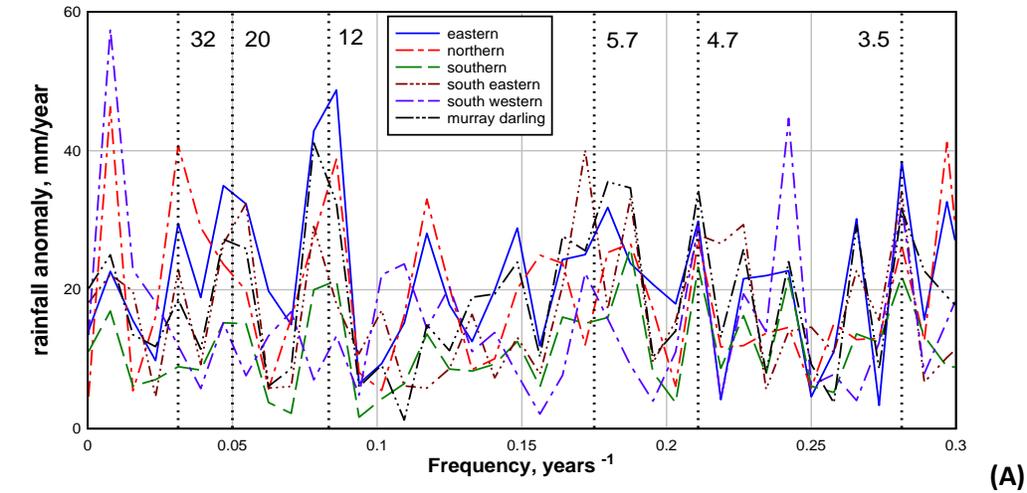

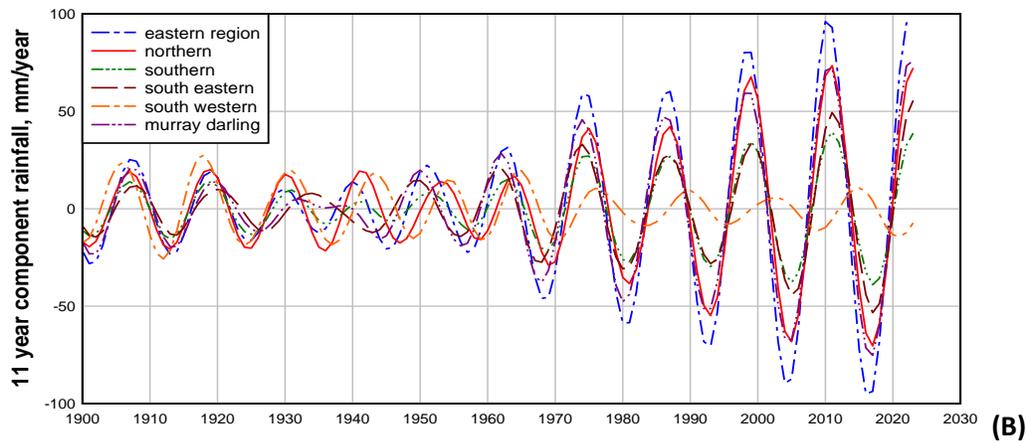

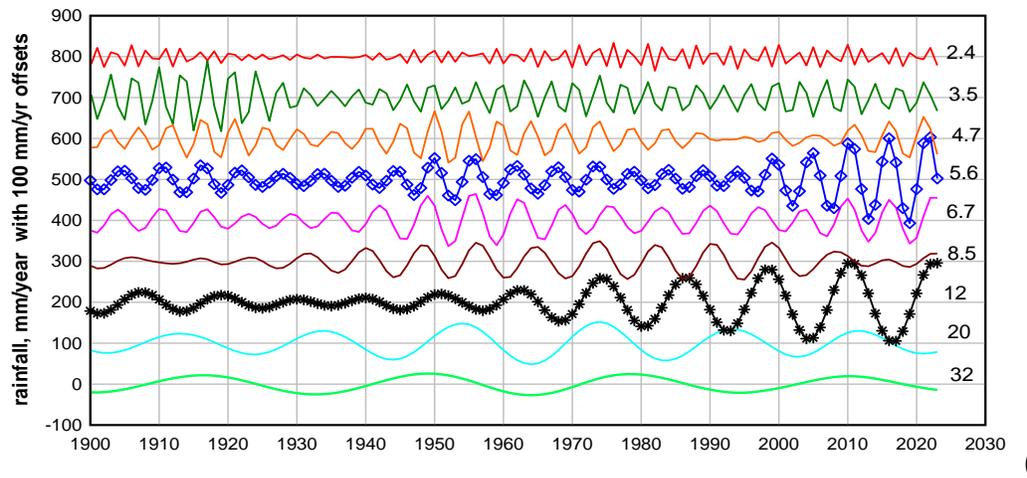

Figure 2. (A). The spectral content of Australian regional rainfall 1900- 2023 (B) The ~ 11 year components of regional rainfall. (C). A frequency/amplitude/time analysis of Eastern Australian rainfall based on components derived by the INF method of band pass filtering.

The question arises as to whether the increasing amplitude of the ~11 year components of rainfall, cloud cover and temperature could be due to the positive trends in rainfall, cloud cover and temperature that are also evident in the data. The trends and means of rainfall, cloud cover and $T_{MAX}$ are available at the BOM site, http://www.bom.gov.au/climate/change/. For Eastern Australia the rainfall trend is +4.1 mm/decade on a mean of 620 m, the cloud cover trend is +0.01 octals/decade on a mean of 3.3 octals, and the $T_{MAX}$ trend is +0.10 C/decade on a mean of 27 C. To examine if the positive trends could be responsible, through some artefact of the INF method of determining components, for the increasing amplitude of the components with time the ~11 year component of $T_{MAX}$ was determined before and after removing the trend, Figure 3B. The increase in amplitude of the ~ 11 year component of TMAX with time is actually slightly greater <u>after</u> any trend is removed. Thus it is clear that the increasing amplitude in the ~11 year cycle components from 1950 onwards is not an artefact of the INF method of finding components.

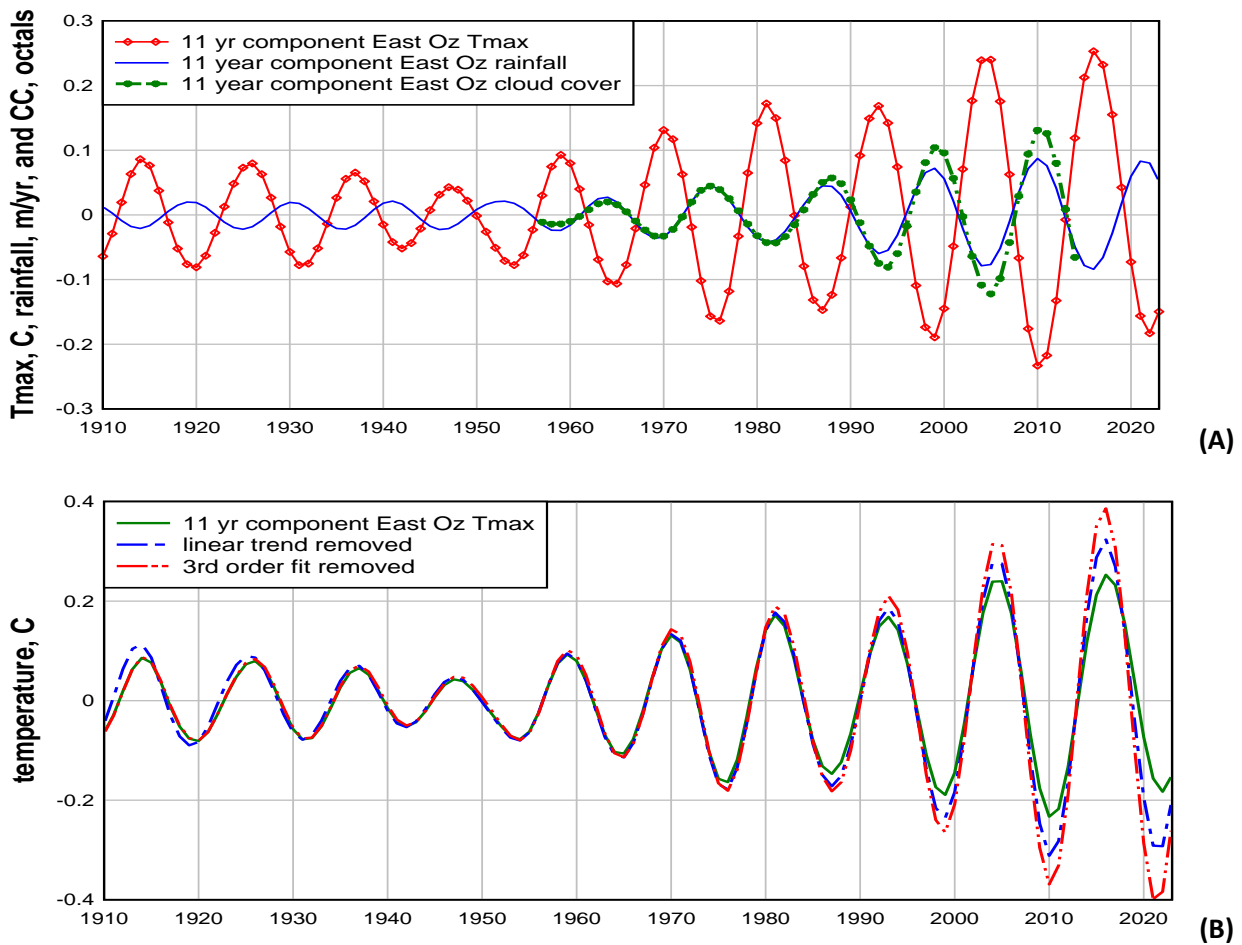

**Figure 3. (A). Shows the decadal components of rainfall, cloud cover and temperature for the Eastern Australian records on the BOM site. The correlations, +ve and –ve, between the components are near exact indicating that, for Eastern Australia, the three climate variables can act as excellent proxies for each other. (B). The decadal component of maximum temperature estimated after two types of trends have been removed from the temperature data. Evidently, the increase in the amplitude of the decadal component is not an artefact of the INF method of obtaining components.**

**3.4. Extending the decadal rainfall record backward in time.** An intriguing aspect of the rainfall observations in the previous sections is that the amplitude of the majority of individual spectral components passes through a minimum at about 1920 -1930, Figure 2C. This effect is especially evident in the decadal component, Figures 2B and 2C. The BOM Climate Change online site provides rainfall data from 1900. However, there are Australian rainfall stations with records in the decades before 1900, (Lavery et al 1992, Lavery et al 1997, Gergis and Ashcroft 2013). The most continuous records are records from lighthouse stations and there are three long rainfall records from Eastern Australia: Cape Moreton Lighthouse (station 040043), Gabo Island Lighthouse (station 084016) and Wilsons Promontory Lighthouse (station 085096). The lighthouse records were averaged over the interval 1873 to 2023 and shown in Figure 4A. As well, a very long, high quality maximum temperature ($T_{MAX}$) record is available from the Sydney Observatory station and, based on the data in Figure 3A showing temperature and rainfall are interchangeable proxies for decadal variability, the $T_{MAX}$ data from this station was used as a proxy for rainfall. The results, normalised on standard deviation, are shown in Figure 4A and show that decadal variability in rainfall in the several decades prior to 1900 is substantial, approaching the level evident in the later decades of the 20th century but not as large as the decadal variability observed after 2000. This result accords with the history of rainfall and drought in Australia, as encapsulated in MacKellers poem of 1908 and with records of historical temperature extremes, Gergis et al (2020).

**4. The ENSO connection to Australian climate at decadal periodicity.**

Climate variability in Australia is mediated via oceanic/atmospheric oscillations principally ENSO SST which is highly correlated/anti-correlated with Australian variation in temperature/rainfall and is similarly broadband. Mechanistic understanding of ENSO is challenging due to the fluctuating and complex nature of the Pacific atmosphere-ocean coupled system with the ENSO following a quasi-periodic process with strong periodicity at 2.5, 3.8, 5, 12-14, 61-75 and 180 years, Bruun et al (2017). Currently, most ENSO models are based on a delayed oscillator model, the variability of the oscillator due to a natural tendency of the periodicity of this type of oscillator to become chaotic or to the effects of internal stochastic forcing (Wang and Picaut 2004, Ruzamaikin (2021). However, most attempts at ENSO prediction are based on extrapolation of a few empirically derived cycles, for example, Bruun et al (2017) suggesting that the current understanding of ENSO variability is limited..

In this article we present the concept of ENSO variability arising from the atmospheric-oceanic oscillation being frequency modulated by the instantaneous frequency of the solar cycle. A solar frequency modulation approach overcomes the conundrum of an essentially monotonic forcing from solar activity generating a broadband phenomena like ENSO because frequency modulating an oscillator intrinsically generates a broadband response in the oscillator, Tibbs et al (1956).

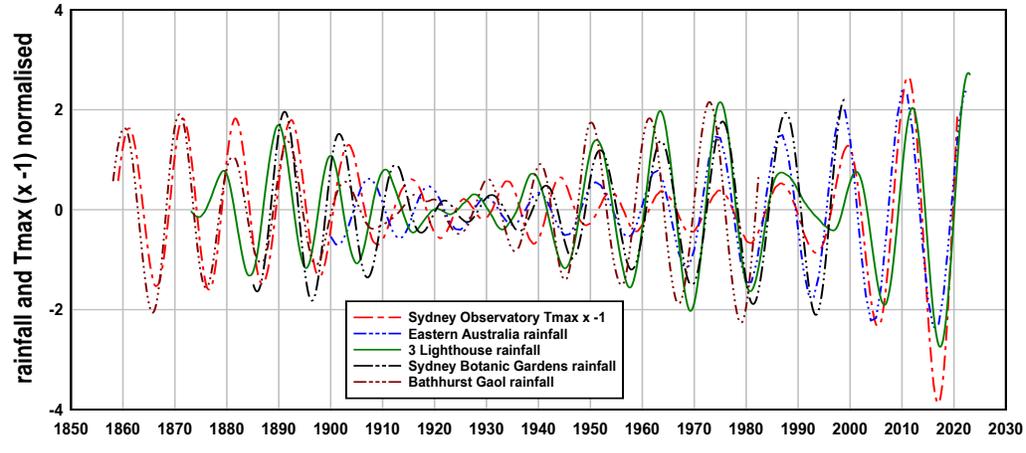

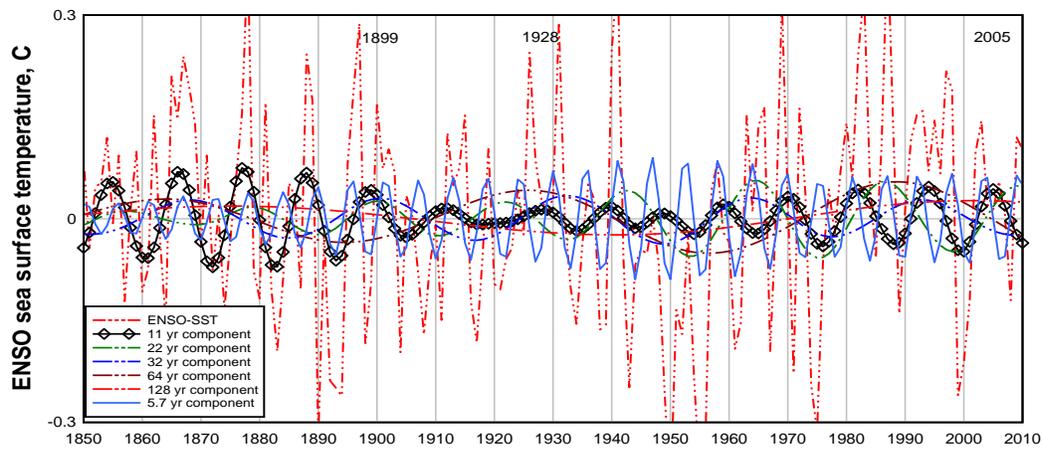

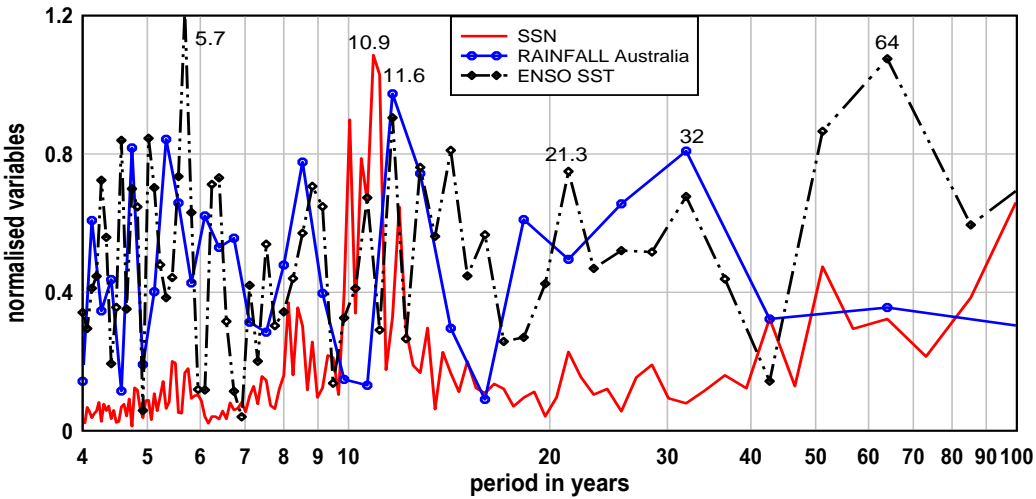

Figure 4. (A).Normalised variations of the rainfall and of $T_{MAX}$ data used as a proxy for rainfall show that the amplitude of the decadal component of rainfall is strong in the late 19$^{th}$ century, decreases to a minimum around 1920 before increasing. (B) ENSO SST and several components including the ~ 11 year component. (C) Periodograms of SSN, ENSO and rainfall Australia illustrate the difference in spectral range between solar and climate variability.

Several research reports study the variation of ENSO in the decadal range, for example, Toure et al (2001).  The ENSO sea surface temperature record 1848 to 2010 may be downloaded at http://research.jisao.washington.edu/data/globalsstenso.  The SST record is shown in Figure 4B as well as several components, including the 11 year component. The amplitude of the 11 year component of ENSO in 1877 is 0.7 C and in 2005 is the amplitude is 0.4 C, a significant decrease. Assuming a direct proportionality between ENSO SST and rainfall is consistent with the result for the decadal component of rainfall, Figure 4A, that indicated high rainfall variability prior to 1900, decreasing to a minimum about 1920, then increasing to a level similar to that prior to 1900 in recent times.

Focusing on the periodicity range between 10 and 13 years in Figure 4C we notice that the periodicity of the solar variable, sunspot number (SSN), is closer to 11 years whereas the periodicity of climate variables, ENSO and rainfall, is closer to 12 years. Figure 5B shows normalised versions of the ~11 year component of ENSO-SST, star symbols, and the ~11 year component of SSN, blue line. Clearly, a large phase shift in the ENSO cycle occurs near 1920 whereas no significant phase shift is evident in the SSN cycle. If a simple estimate of average period is made by dividing the time interval between the first and last positive peaks of each cycle the result is 10.8 years for the SSN cycle and 11.5 years for the ENSO-SST cycle, a difference of about 6% in period. Thus, the phase shift in the ENSO cycle near 1920 explains why different, but approximately decadal, periodicities are obtained for solar and climate variability. Past work that accessed sea surface temperature records from 1900 to present, for example White et al (1997), reported a positive correlation between solar activity and temperature. However, the ENSO decadal cycle is essentially in anti-phase with the SSN decadal variation before 1920 and is essentially in phase with the decadal SSN after 1920 and this precludes a direct connection between solar cycle amplitude and climate variability.

**5. The frequency variation of the solar cycle.**

The Sun provides the principal energy input into the Earth system and solar variability represents a significant external climate forcing, Brehm et al (2021).  Historically, solar activity has been measured by counting sunspots, Hathaway (2015). Accurate sunspot counts have been available from about 1750, and can be used to find the change in solar cycle length or period as a function of time, https://www.sidc.be/SILSO/cyclesmm.

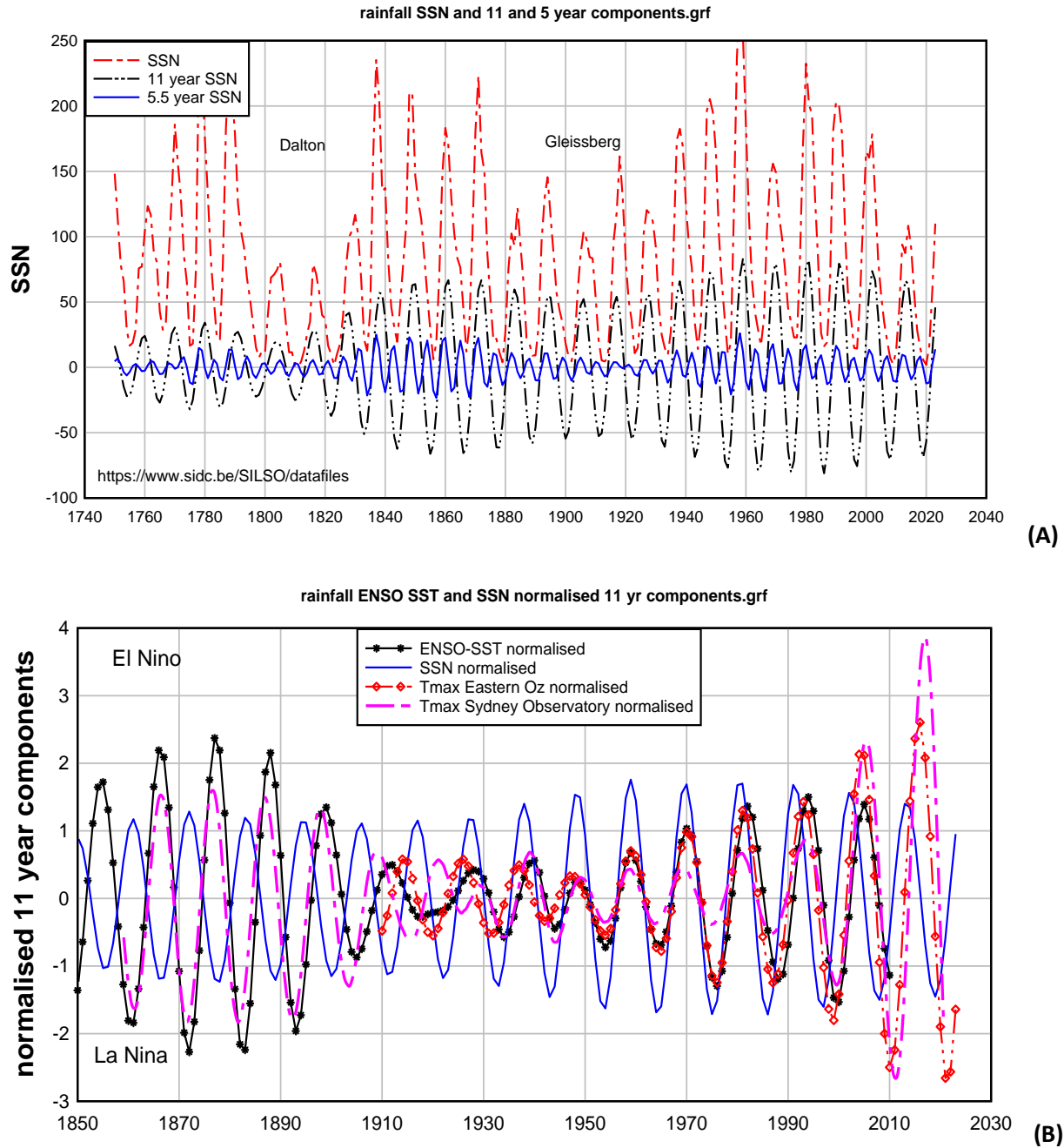

**Figure 5. (A)** The directly observed sunspot number, SSN, 1748 to 2022. Also shown the ~11 year component and the ~ 5.5 year of the sunspot number, SSN. The Dalton and Gleissberg grand minima in SSN are labelled. **(B)** A comparison of normalised ~11 year components of ENSO sea surface temperature, SSN, $T_{MAX}$ Eastern Australia and $T_{MAX}$ Sydney Observatory. Note that the ~11 year components of ENSO SST and $T_{MAX}$ are consistently in-phase over the entire record and are approximately in-phase with the ~11 year component of SSN from 1930 to 2023. However, the ~11 year components of the climate variables ENSO SST and $T_{MAX}$ are in anti-phase with the ~11 component of SSN in the second half of the 19$^{th}$ century. This precludes a direct relationship between climate variability and solar variability.

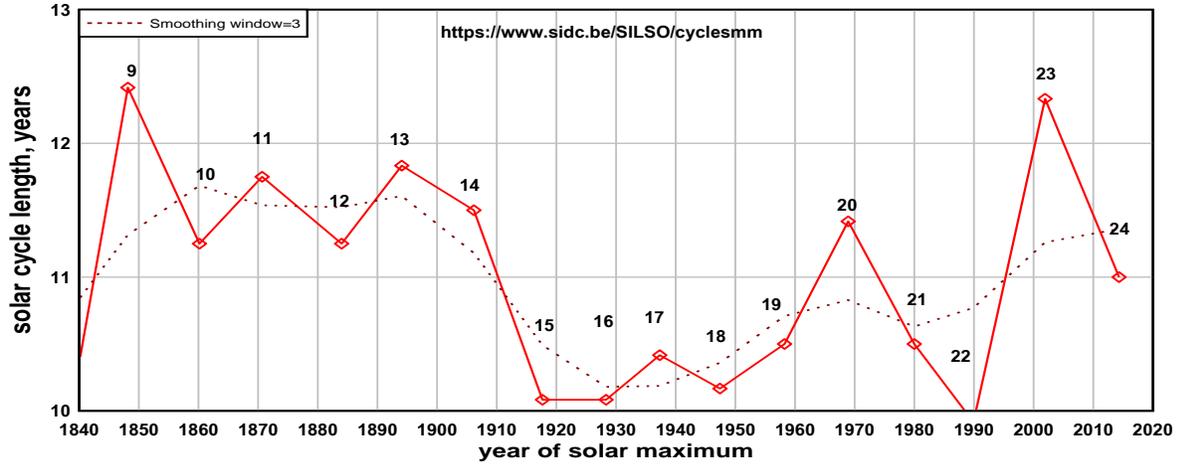

(A)

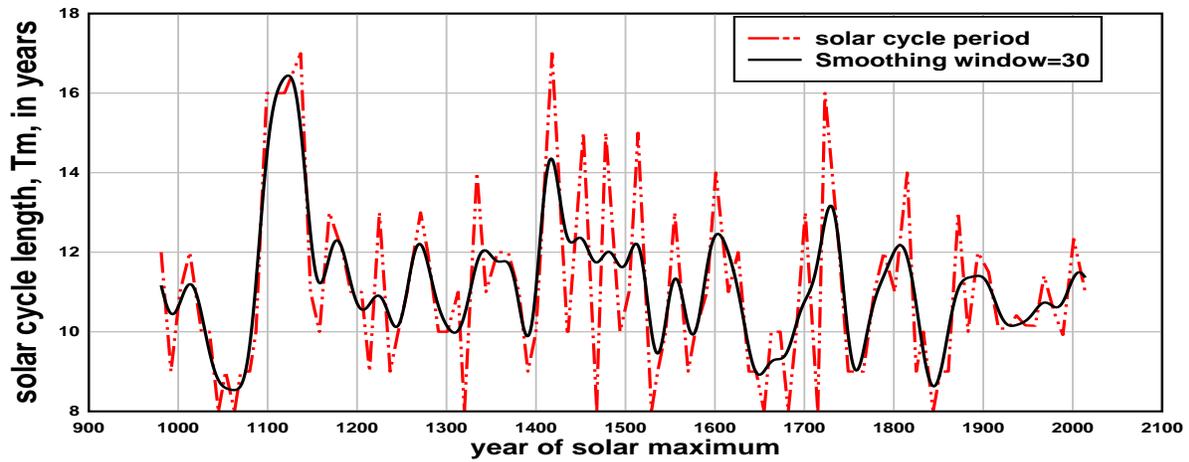

(B)

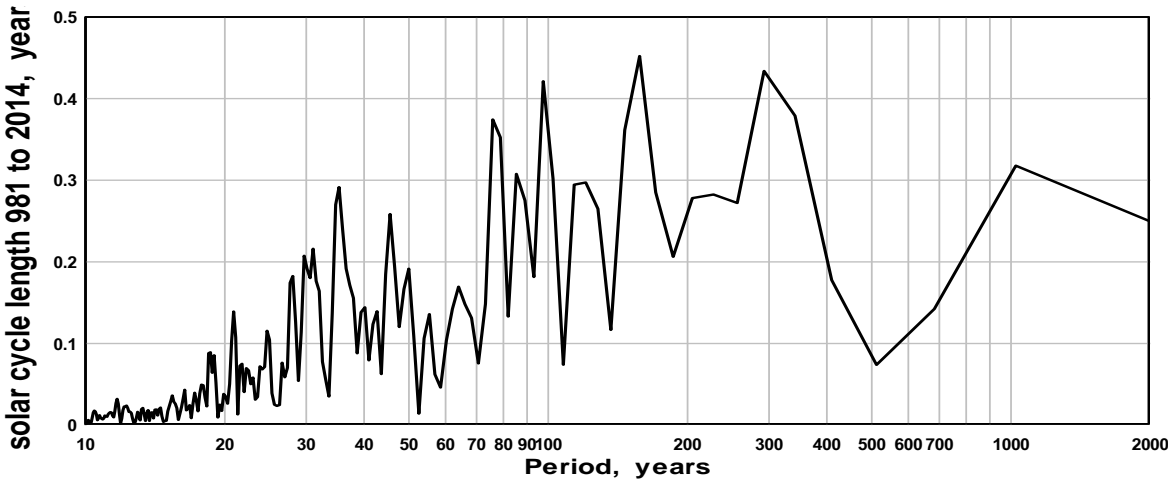

(C)

Figure 6. (A) The variation of solar cycle length or period, P, with time during 1840 to 2014 exhibits a long period, P ~ 100 years with a shorter period, P ~ 30 years evident towards the end of the record, 1960 – 2014. (B) With the reconstructed solar length variation added the record extends from 981 to 2014. (C) The periodogram of solar cycle length indicates periodicities extending from ~20 years to ~1000 years.

The observed solar cycle length as a function of year of solar maximum is shown in Figure 6A. For times prior to 1750 proxy records of solar activity are required. Galactic cosmic rays generate $^{14}$C in the atmosphere and, as the cosmic ray flux varies with solar activity, there is a small change in the $^{14}$C content of the atmosphere between solar maximum and minimum. As a result the $^{14}$C stored in the annual growth rings of trees varies and measuring the $^{14}$C content in tree ring by mass spectrometry allows the annual strength and timing of the solar cycle to be measured and allows the solar cycle length to be found back to year 981, (Brehm et al 2021, Usoskin et al 2021). The data on solar cycle length, 976 to 1894, provided by Usoskin et al (2021) was downloaded at https://www.ngdc.noaa.gov/stp/solar/solardataservices.html and along with the recent solar cycle lengths is shown in Figure 6B. Evidently the solar cycle period varies in a rather complicated way between 8 and 17 years. The periodogram of solar cycle periodicity, P, shown in Figure 6C, indicates the variability in solar cycle length has components ranging from about 20 years up to about 1000 years with the strongest components at periods between 30 to 300 years. Band pass filtering indicates that the longer period variations, P ~100 years, tend to persist over the entire millennium whereas shorter period variation, P ~ 30 years, is dominant only for relatively short intervals, for example, 1400 to 1500 and 1650 to 1750, Figure 7. Since these intervals coincide with the Sporer and Maunder grand solar minima respectively the shorter period variation may be an artefact due to the $^{14}$C level falling below the noise level, (Brehm et al 2021, Usoskin et al 2021).

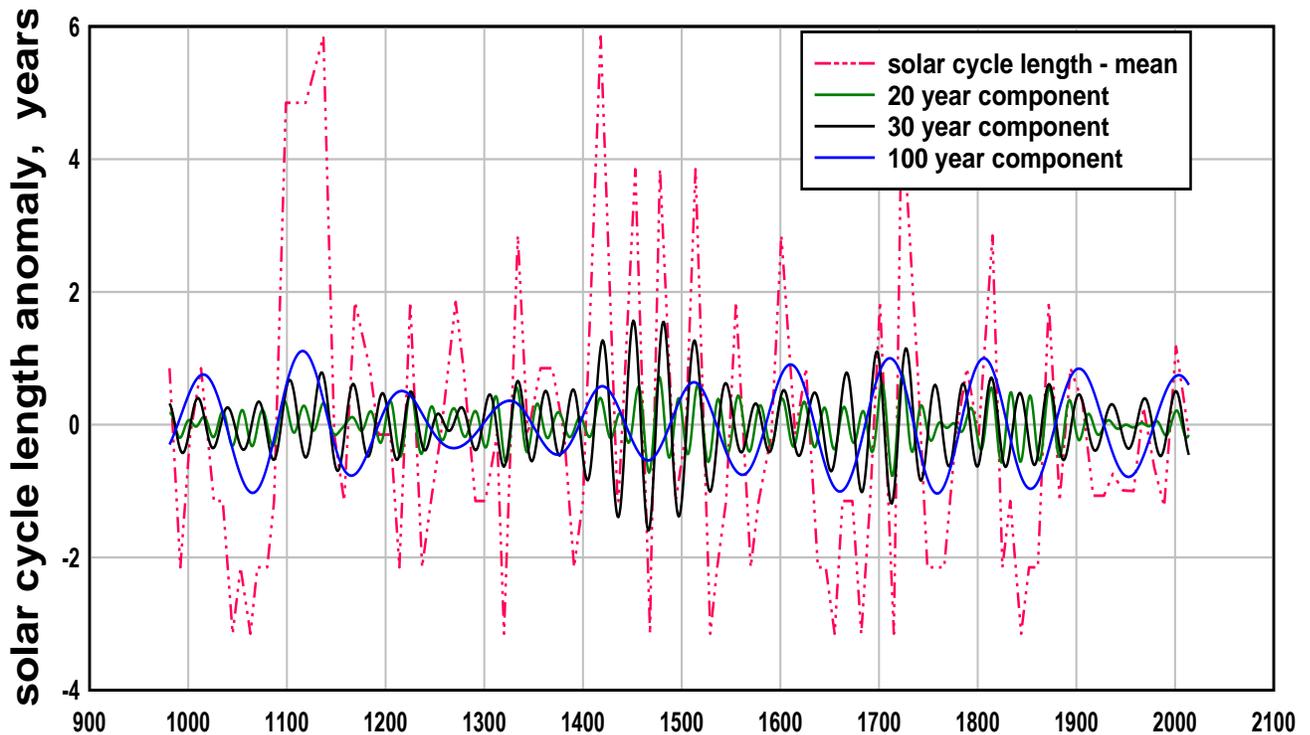

**Figure 7. Band pass filtered components of the solar cycle length anomaly indicate dominance of a low frequency component, P ~ 100 years, from 1000 to 1200 and from 1800 to 2000 and evidence of dominance of a high frequency component, P ~ 30 years, from 1400 – 1500 and from 1650 – 1750.**

## 6. Modelling a frequency modulated El Nino Southern Oscillation.

**6.1 Theoretical model.** Various delayed oscillator models of oceanic/atmospheric oscillations have been proposed, that, supposedly, develop a chaotic variability internally, Wang and Picaut 2004, Bruun et al 2017), or are forced by stochastic weather processes, Liu and Zhang (2022), or are forced by the solar cycle, (White et al 2003, White and Liu 2008). However, none of the models relate to the frequency modulation of ENSO. It is known that the frequency of strong El Nino events varies on a centenary time scale, Enfield and Cid S (1991). Thus the question is not whether ENSO is frequency modulated but why and how ENSO is frequency modulated? Here, we assume that ENSO is a simple harmonic oscillator, (SHO), and is forced by a solar cycle of varying frequency. The response of a SHO with natural angular frequency $\omega_0$ and damping factor, $\gamma$, to constant frequency forcing varying as $\sin(\omega t)$ is derived in most basic physics and engineering texts. The response, $x(t)$, is given by

$$x(t) = (k/\omega Z)\sin(\omega t + \phi) \tag{1}$$

where k is a constant and

$$Z = ((2\omega_0\gamma)^2 + (\omega_0^2 - \omega^2)^2/\omega^2)^{1/2}, \tag{2}$$

$$\tan(\phi) = 2\omega\omega_0\gamma/(\omega^2 - \omega_0^2), \text{ and} \tag{3}$$

$$\phi = \tan^{-1}(\tan(\phi)) - \pi/2 \tag{4}$$

Assuming ENSO behaves like a SHO then if ENSO was forced by a solar cycle of period 11 years and angular frequency $\omega = 2\pi/11$ year$^{-1}$ the ENSO response, $x(t)$, would be sinusoidal at period 11 years and at some constant phase difference, $\phi$, relative to an 11 year solar cycle. However, as indicated in the previous section the instantaneous period, $T_M(t)$, of the solar cycle varies with time and $\omega$ in the above equations should be replaced by

$$\omega_M(t) = 2\pi/T_M(t) \tag{5}$$

where $T_M(t)$, the instantaneous solar cycle period varies as indicated in Figures 6 and 7. In this scenario the variation of ENSO sea surface temperature is now frequency modulated by the solar cycle. Frequency modulation has a profound effect on the time variation and the frequency variation of the modulated oscillator. The principal changes are that the oscillator output varies in frequency and the spectral content of the oscillations becomes broad band, i.e. displaying a wide range of frequencies, (Tibbs et al 1956, Faruque 2017). If the period of the frequency modulation is variable, as is evident from Figures 6 and 7, the time variation and spectral content of the ENSO model becomes very complex. However, there are times when one solar modulation period becomes dominant. For example, referring to Figure 7, a solar cycle modulation period of ~ 30 years is dominant during the 1400 to 1500 interval and a solar cycle modulation period of ~ 100 years is dominant during the 1800 to 2000 interval.

If one modulation periodicity, P, is dominant, for example the P = 100 year solar cycle modulation period in the 1800 to 2000 interval evident in Figure 7, the frequency modulation term $\omega_M(t)$ in equation 5 can be approximated by a relation of the form

$$\omega_M(t) = 2\pi/[T_{M0}(1 + A\cos(2\pi(t - t_0)/P))] \tag{6}$$

Where $T_{M0}$ is the average long term period of the solar cycle in this interval, $A$ is expressed as $\Delta T_{M0}/T_{M0}$ where $\Delta T_{M0}$ is the maximum deviation from $T_{M0}$ in this interval and $t_0$ is a year when the modulation at period P is at maximum. For example, referring to Figure 7 and the P = 100 year component, $T_{M0}$ is about 11 years, $\Delta T_{M0}$ is about 1 year so $A \sim 1/11 = 0.09$, and $t_0 \sim 2000$. In communication engineering terminology this form of modulation is referred to as single-tone frequency modulation and, in this case, the time variation of the response can be expressed analytically as an infinite series of Bessel functions, Tibbs et al (1956). An and Jin (2011) obtained similar expressions as the equations above for frequency modulation of a simple harmonic model of ENSO by the annual cycle.

**6.2. Fitting the ENSO model to observations.** Here, we attempt to fit the model as in equations 1 to 6 to the observed decadal component of ENSO, Figures 4B and 5B. There are two sets of parameters to be considered. One set concerns the natural angular frequency, $\omega_0$, and the damping factor, $\gamma$, of ENSO. The other set concerns the long term period of the solar cycle, $T_{M0}$, the dominant modulation period, P, in the time interval of interest, the modulation factor $A$, and the phase of the modulation or the time, $t_0$, when the modulation of period P is at maximum. This set of parameters is moderately well defined by the observed instantaneous periodicity of the solar cycle, Figures 6A, 6B and Figure 7. We know, from Figures 6A and 6B, that $T_{M0}$ is close to 10.8 years, and $A = \Delta T_{M0}/T_{M0}$, is about 0.09. From Figure 7, P is about 100 years and $t_0$ is about year 2000. These estimates can be used as starting values for the fit. The two parameters concerning the natural oscillation of ENSO, the natural period, $T_0$, and the damping factor, $\gamma$, are less clearly defined. Here we take $T_0$ to be 11.8 years as this is close to the most prominent periodicity observed in Australian rainfall spectra, Figure 2A. Also, 11.8 years is close to the average time interval between La Nina events, 12.1 years during 1900 – 1999, that is included in global climate modelling studies, Geng et al (2023). The damping factor, $\gamma$, of the natural oscillation is taken initially as $\gamma = 0.1$. The parameters were adjusted while the 11 year component of the ENSO model was compared with the 11 year component of the observed ENSO SST variation, Figure 4B. The result of the fit is shown in Figure 9A. However, before discussing the fit, we outline some details of the ENSO model. The fitted parameters relating to the solar cycle length variation were, a single tone modulation of period P = 100 years, long term average solar cycle period, $T_{M0}$ = 10.8 years, modulation factor $A$ = 0.02, and $t_0$ = 2010. The solar cycle period variation, P = 100 years, used in the model is shown in Figure 8A and is similar in phase and in period but is about one quarter the amplitude of the observed solar cycle variation, Figures 6B and 7. The fitted parameters relating to the natural oscillation of ENSO are, natural period, $T_0$ = 11.8 years, and damping factor $\gamma$ = 0.1. The amplitude and phase response of the modelled ENSO to direct forcing as a function of period, T, of forcing are shown in Figure 8B. Also shown in Figure 8B are a solid reference line marking the long term average solar periodicity, $T_{M0}$ = 10.8 years, and two dotted reference lines marking the maximum positive and negative deviation from $T_{M0}$ over the 100 year modulation cycle. The 100 year period variation in solar cycle period results in a variation in amplitude and phase of the ENSO model response within the limit of the dotted reference lines as the solar cycle length varies as in Figure 8A. The frequency modulated response of the ENSO model is shown in Figure 8C for the time interval 1750 to 2040 along with the 11 year component of the ENSO model response.

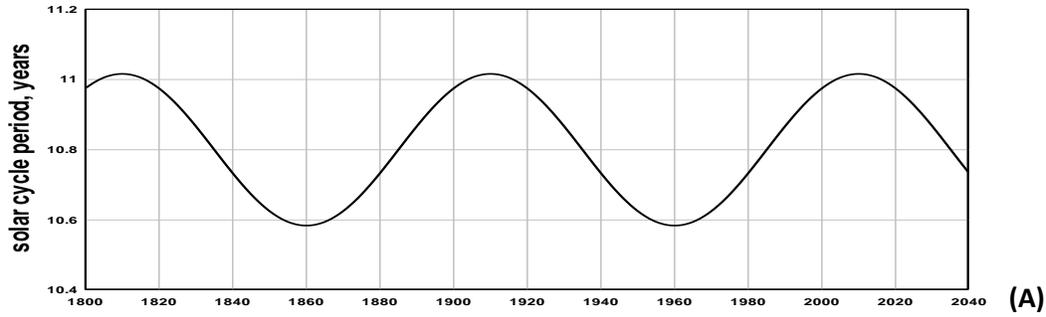
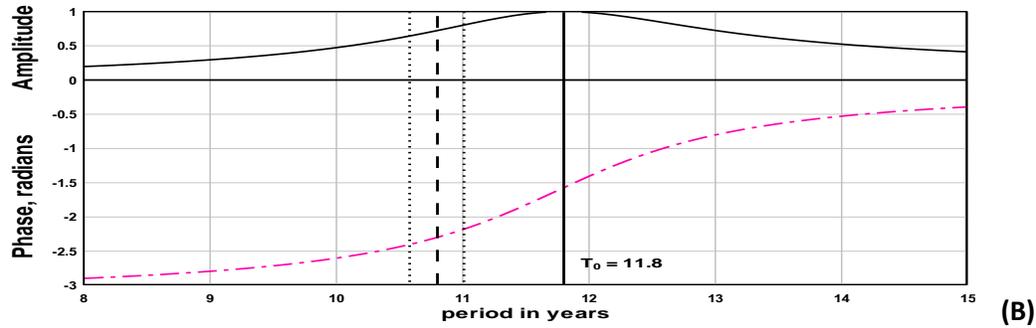
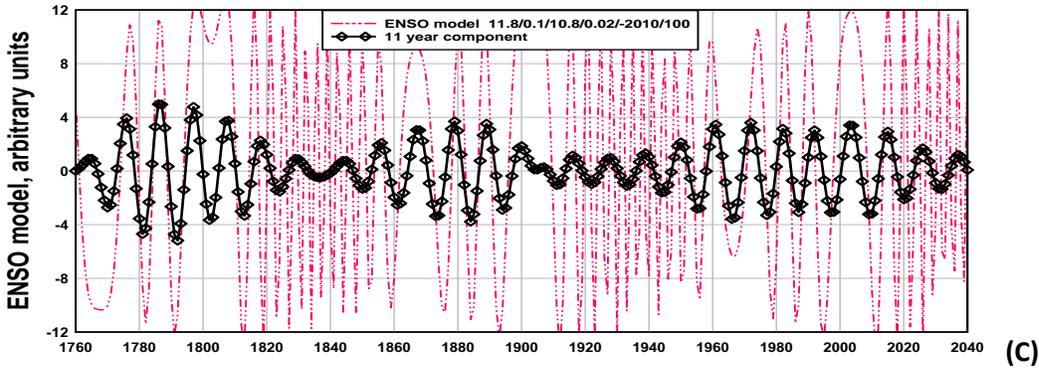
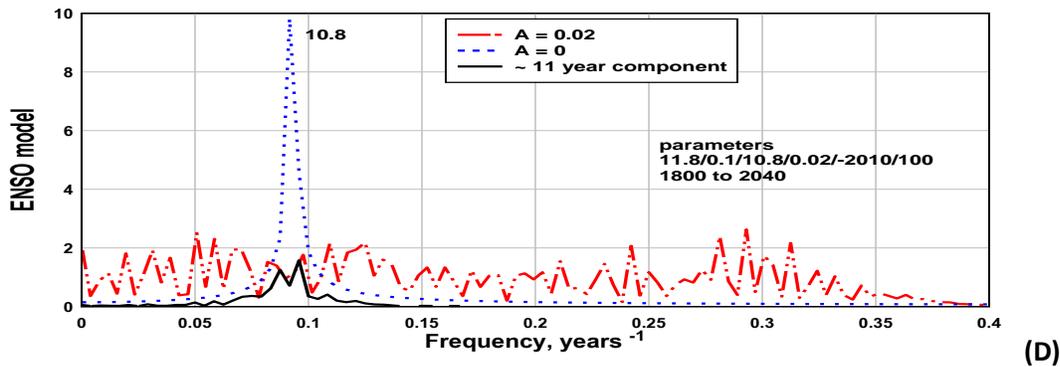

Figure 8. (A) The single-tone solar cycle length variation, P = 100 years, forcing the model ENSO. (B) The natural amplitude and phase response as a function of period for the model ENSO. The frequency forcing of the oscillator is centred on 10.8 years, full reference line, and the 100 year period sinusoidal modulation varies between the dotted lines. (C) The resulting ENSO model response versus time and the 11 year component of the response. (D) The frequency spectrums of ENSO model response. Blue curve: zero frequency modulation, red curve: frequency modulation factor, 0.02, black curve the spectral content of the 11 year component.

Comparison of the model ENSO response in Figure 8C with the time variation of solar period in Figure 8A shows that when the solar cycle length is decreasing, e.g. during the 1820 to 1850 and 1920 to 1950 intervals, the ENSO response is changing at high frequency and when the solar cycle length is increasing, e.g. during the 1870 to 1900 interval, the ENSO response is changing at low frequency. It is known that the observed ENSO has intervals of fast change and slow change, Cane (2005). This is obvious when the records of ENSO sea surface temperature and Southern Oscillation sea level pressure are presented as three consecutive segments in time, as in Cane (2005). Thus the time variation of the ENSO model replicates this aspect of observed ENSO. The Fourier spectra of the frequency modulated ENSO model variation, (A = 0.02), the non frequency modulated ENSO model variation, (A = 0), and the 11 year component of the ENSO model variation are shown in Figure 8D. The spectrums illustrate the principal feature of frequency modulation of an oscillator: a flat, broad band spectrum results from solar forcing when the frequency of the solar forcing varies, in this case with a 100 year period. The spectrum of the ~ 11 year component of the model ENSO variation is also shown and indicates that the time variation of the 11 year component can be regarded as the result of interference between two components, one of period 10.4 years, the other of period 11.4 years, with a beat period ~120 years and average period 10.9 years.

The 11 year period component of the ENSO model in Figure 8C is compared with the 11 year component of observed ENSO sea surface temperature in Figure 9A. The correlation is remarkably good, conveying the major features of the 11 year component in ENSO SST: the strong variation in the late 19th and late 20th centuries, the relatively much weaker variation in the early 20th century and the $\pi$ phase shift relative to the average period from one long term beat to the next. Compare with Figure 5B where the $\pi$ phase shift between the observed 11 year component of ENSO SST and the sunspot number variation is evident. Thus the model replicates most of the observed ENSO 11 year component features. To extend the comparison of the ENSO model over longer time intervals it is necessary to utilise reconstructed paleo-climate records. The 11 year period component of the ENSO model in Figure 8C is compared with the 11 year component of reconstructed Northern Australia Standardized Precipitation – Evaporation Index, Allen et al (2020), in Figure 9B. Note that the ENSO model variation has been inverted to facilitate the comparison with the precipitation index in Figure 9B. Again, the correlation with between the ENSO model and this 246 year paleo-climate record is remarkably good.

Fits to other components of ENSO provide similar results with only minor variations in the fitting parameters. However, as we are interested in providing supporting evidence that the model can be applied successfully to a wide range of climate data, we consider below the model comparison with other climate variables.

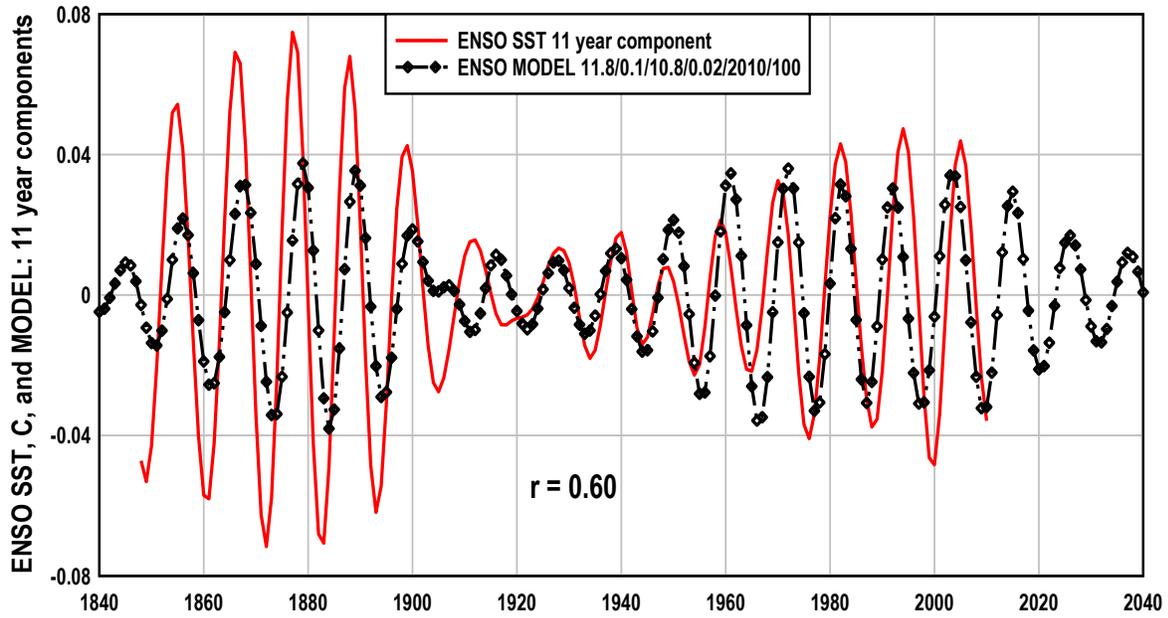

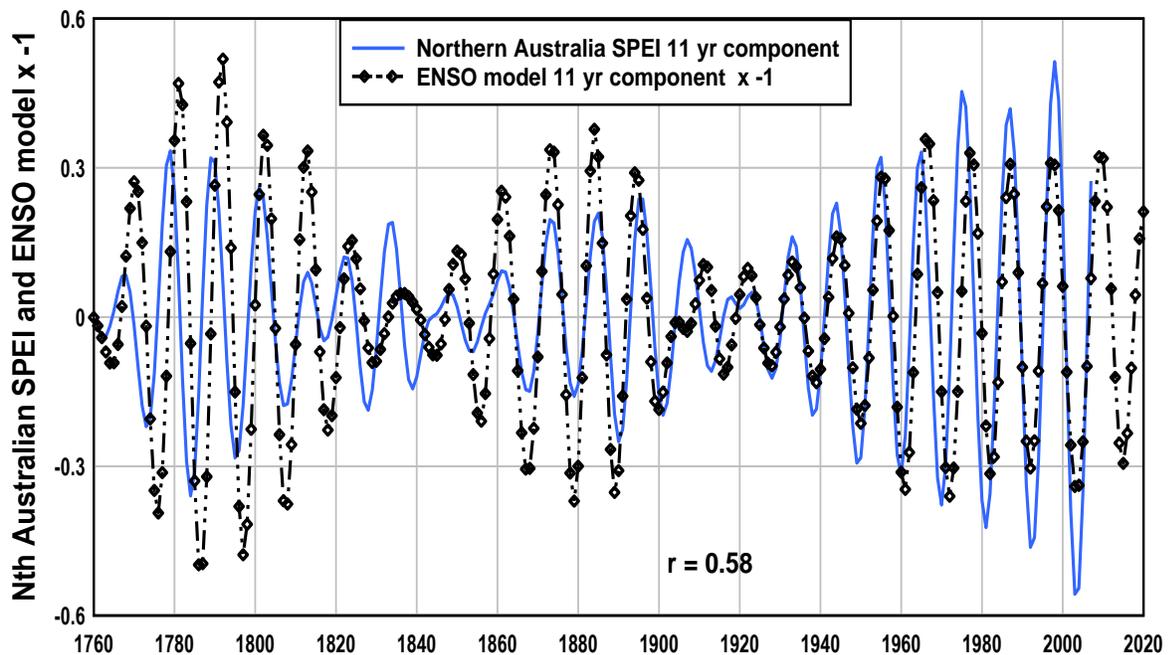

Figure 9. (A) A comparison of the 11 year component in the observed ENSO SST and the 11 year component in the modelled ENSO, P = 100 years. The sequence of values of parameters used in the model are 11.8/0.1/10.8/0.02/2010/100 corresponding, in sequence, to the following parameters in equations 1 – 6: $T_0$, $\gamma$, $T_{M0}$, A, $t_0$ and P. (B) An inverted version of the 11 year component of the ENSO model compared with the 11 year component of the reconstructed North Australia SPE index, an index of rainfall, Allen et al (2020).

## 7. Supporting evidence for frequency modulation of climate variability by solar cycle periodicity.

**7.1 The effect of solar cycle modulation in Australian rainfall data.** Figure 8C shows the time variation of the ENSO model when the imposed solar cycle modulation period P = 100 years. The evident feature is the cyclic change of the ENSO from slow variability to faster variability. For example, fast variability predominates from about 1900 to 1960 and slower variability predominates from 1960 to 2020. This change from faster to slower variability is expected to be evident in the spectra of Australian rainfall and temperature data as these climate variables are highly correlated with ENSO. Figure 10A shows the frequency spectra of regional Australian rainfall for the interval 1900 – 1961, the first half of the BOM rainfall data record. As expected, the variability occurs predominantly in higher frequency components with periods between 2 and 6 years. Figure 10B shows the regional spectra for the latter half of the rainfall record, 1962 to 2023. Now the variability is predominantly in lower frequency components with periods between 5 and 40 years. The observed spectrums are consistent with ENSO the model time variation.

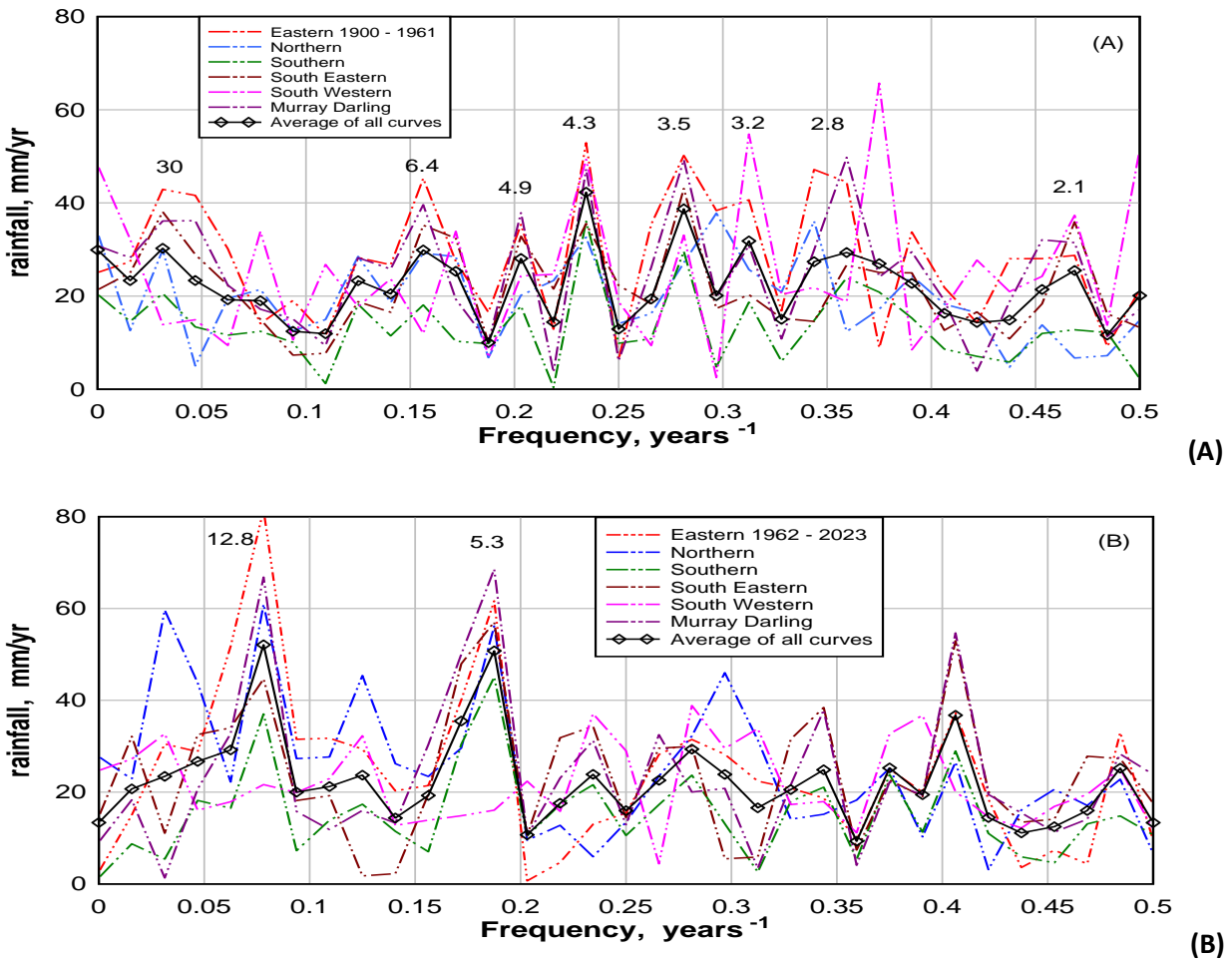

**Figure 10.** (A) The spectra of Australian regional rainfall during 1900 to 1960 is due, mainly, to short term, 2 - 6 year variability, whereas, (B) the spectra during 1961 to 2023 is due, mainly to longer term variability.

**7.2 The double Fourier transform as an empirical link between solar and climate variability.** The time variation of a frequency modulated variable is relatively simple when the modulation is single-tone as was the case for the modelling example in the previous section, Figure 8, where modulation was at period 100 years. In the case of single-tone frequency modulation the spectral content of the oscillator response is expected to be broadband but relatively simple. Analytical details of the spectrum of a frequency modulated variable are given in communication engineering texts, for example Tibbs et al (1956), and are provided in illustrated introductions to frequency modulation, for example Faruque (2017). The spectra is comprised of components at frequencies given by $f_{M0}$ +/- $nf_P$ where $f_{M0}$ = $1/T_{M0}$ and $f_P$ = $1/P$ and n = 1, 2, 3, 4 .... etc. Thus peaks on either side of $f_{M0}$ are spaced at intervals $f_P$. So for the case where P = 100 years the spectral peaks are at frequencies 1/10.8 +/- n/100 or 0.0926 +/- n0.01 years$^{-1}$. Here we illustrate the expected spectrum of modelled ENSO empirically by applying a FFT to the ENSO model time variation with the time variation extending over 1000 years, 1500 to 2500, Figure 11A, to achieve higher resolution in the resulting frequency spectrum, Figure 11B. The obvious feature of the frequency modulation spectrum in Figure 11B is the occurrence of a sequence of peaks at frequency separation $f_P$ = $1/P$ = 0.01 years$^{-1}$. The regular spacing of spectral peaks is most clearly evident towards the higher frequency end of the spectrum. At lower frequencies the spectral peaks are less evenly spaced and some occur at spacing $f_P/2$, $f_P/3$ ... etc. It is obvious that if a second Fourier transform is made of the first Fourier spectrum the result is a transform into the time domain. A peak will be evident most strongly at period $T_P$ = $1/f_P$, and there will also be peaks at periods = $2/f_P$, $3/f_P$ etc, see Figure 12A. This modelling result indicates that the solar cycle modulation period should be recoverable from climate data by using this double Fourier transform method provided there is a dominant solar cycle modulation period. When more than one solar modulation period is present the spectral content of the modulated variable becomes very complex. The case of frequency modulation comprising two sinusoidal components of period P1 and P2 is discussed in Tibbs et al (1956). Whereas with single tone modulation the spectral peaks occur at $f_{M0}$ +/- $nf_P$ with spacing $f_P$, with dual tone modulation spectral peaks occur at $f_{M0}$ +/- ($nf_{P1}$ +/- $mf_{P2}$) and are at multiple frequency spacing, ($nf_{P1}$ +/- $mf_{P2}$), where n, m = 0, 1, 2, 3 etc. In this case the double Fourier transform of the modulated variable will yield multiple periods. Figure 11B is an example of the recovered periods for dual-tone modulation.

This method of recovering solar cycle length periodicity, called here the double Fourier transform method, can, in principle, provide a link from climate records back to the known solar cycle periodicity. We examine a few long instrumental climate records in the next section to test this concept. Another factor in recovering solar cycle length modulation periods in climate data is that, generally, the solar modulation is not single-tone, i.e. there are, at all times, two or more solar cycle modulations present, see Figure 7.

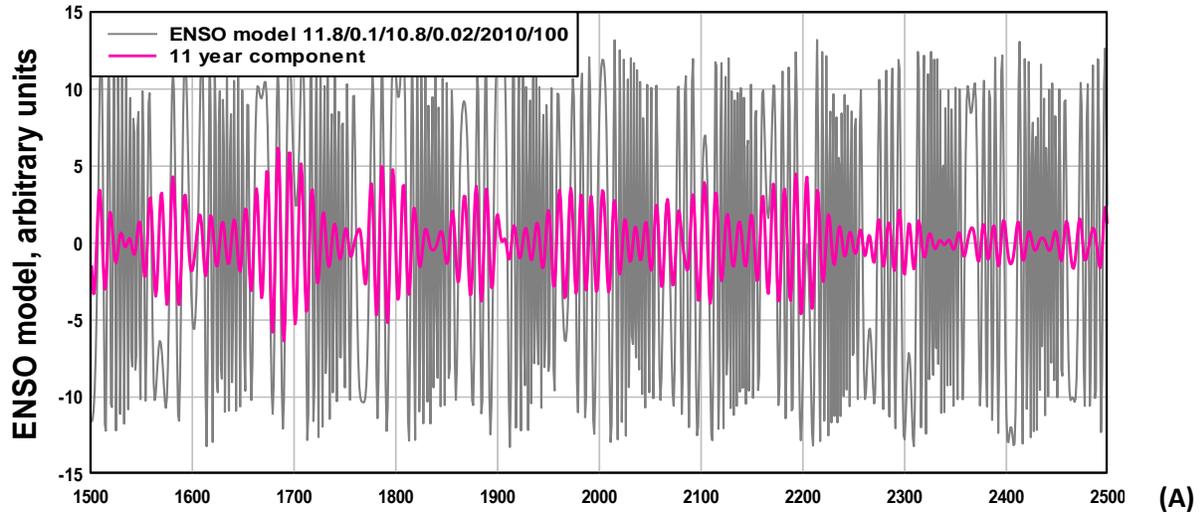

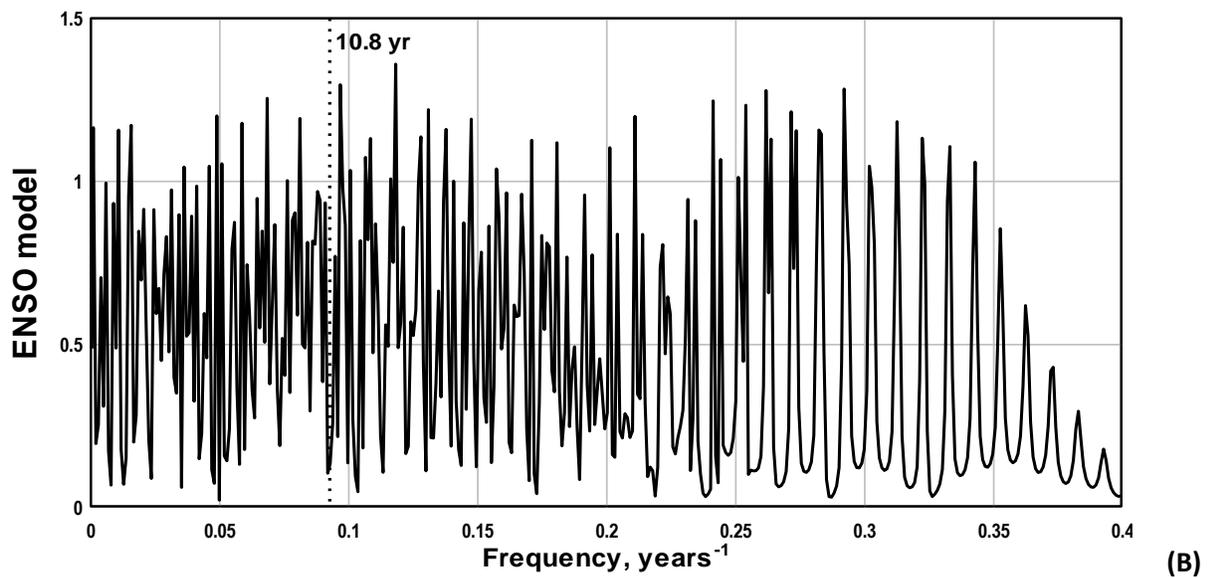

Figure 11. (A) The model ENSO time variation when the periodicity in the solar cycle length variation is single tone of period P = 100 years. (B) The first Fourier transform of the ENSO model variation gives a spectrum where the peak spacing is mainly at 1/P = 0.01 years$^{-1}$. The even spacing of peaks is most evident towards the higher frequency end of the spectrum with finer spacing, 1/2P, 1/3P, evident towards the lower frequency end of the spectrum.

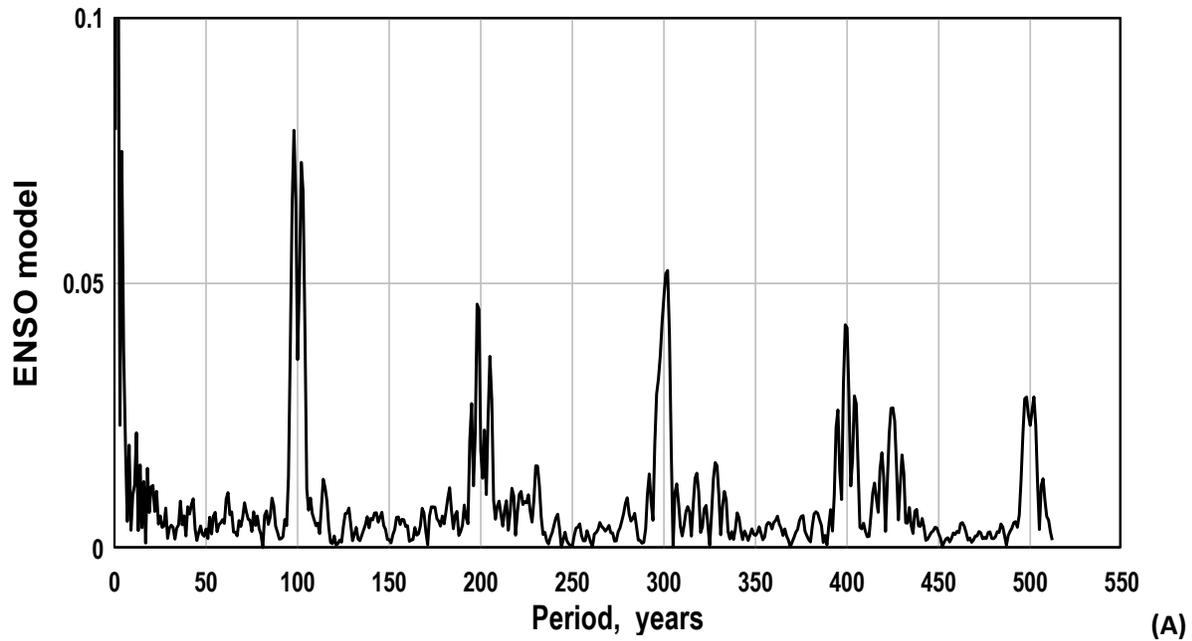

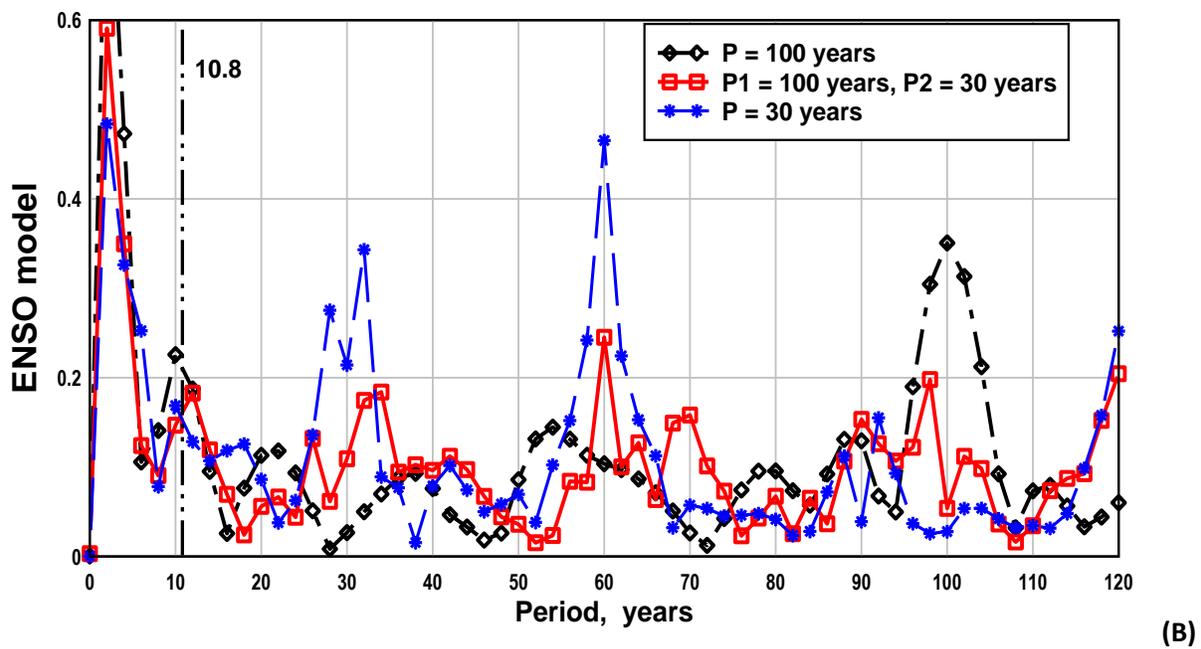

Figure 12. Illustrates the recovery of solar cycle modulation periodicity using the double Fourier transform method. (A) Shows the second Fourier transform of the ENSO model spectrum in Figure 11B. The solar cycle length period, P = 100 years is recovered as well as sub harmonics. (B) Illustrates solar cycle length periodicity when the ENSO model time variation extends from 1800 to 2040, i.e. 240 years. In the two single-tone modulation cases cycle length periods of P = 100 and P = 30 years are readily recovered. However, in the case of equal strength dual-tone modulation, recovery of the shorter modulation period, P = 30 years, is more apparent than recovery of the longer modulation period, P = 100 year

To examine the effect of multiple period modulation we generate the ENSO model time variation, again between years 1500 and 2500, but with two modulations present, one with P = 100 years and A = 0.005 and one with P = 30 years and A = 0.005. Thus the modulation strength $\Delta T_M$ at P = 30 years equals the modulation strength $\Delta T_M$ at P = 100 years. Now the model time variation is very complex, with the appearance of chaotic variability, (not shown here), and the spectral content of the first Fourier transform variation is more diffuse. However, in the double Fourier transform the 30 year modulation period is recovered, Figure 12B, while the 100 year modulation not recovered. Thus, the expectation is that a solar modulation period should be recoverable from climate records provided the time interval selected, by reference to the solar cycle length anomaly, Figure 7, is an interval when one solar cycle length modulation period is dominant. We assess this concept in the following sections.

**7.3. Solar frequency modulation periods in climate records.** Here we apply the double Fourier transform method to rainfall, temperature and oceanic/atmospheric oscillation records.

**7.3.1 Australian rainfall and temperature records.** Referring to the solar cycle length anomaly of Figure 7, during the time interval, 1900 to 2023, of the Australian regional rainfall data, solar cycle length modulation periods of 30 and 100 years are strong. In the case of dual-tone modulation the spacing's, $\Delta f_{n,m}$, of the peaks in the first Fourier transform are given by

$$\Delta f_{n,m} = (nf_{P1} +/- mf_{P2}) \tag{7}$$

and the recovered periods in the second Fourier transform are expected at $1/\Delta f_{n,m}$. For n, m = 0, 1, 2 the expected periods are 11.5, 13.0, 15.0, 17.6, 18.7, 21.4, 23.0, 30.0, 42.8, 50.0, 75.0 and 100.0 years. For example, $(1/30 - 2/100)^{-1} = 75$ and the order is n + m = 3, with higher order peaks in the double transform usually less strong than lower order peaks. Peaks at sub harmonics of the modulation periodicity are also expected as indicated in Figure 12A. For example, the 30 year modulation period is expected to be accompanied by peaks at 60, 90 and 120 years. The frequency spectrums of the six regional Australian rainfall records were shown earlier in Figure 2A. The record length, 123 years is too short to recover 100 year modulation periodicity in the double Fourier transform. However, when the padding of the rainfall data is extended from $2^7 = 128$ years to $2^9 = 512$ years and the double Fourier transform applied periods up to 256 years are recoverable although periodicity > 120 years is negligible, Figure 13A. Strong modulation periodicity is centred on 10, 15, 30, 40, 60, 75 and 100 years; i.e. apart from the sub harmonics of 30 years at 60, 90 and 120 years the other periods are predicted by the dual tone equation 7. Most of the aforementioned peaks are also evident in Figure 13B where the regional Australian $T_{MAX}$ data has been treated by the same double Fourier method. This provides strong support for the frequency modulation concept. However, the ten periods just mentioned are not the only recovered periods present. Examples are the strong peaks at 10 years and at 27 years. The peak at 10 years corresponds to the higher order n = 3, m = 0 case and the peak at 27 corresponds to the even higher order n = 2, m = 3 case. Nearly all the periods relating to peaks in Figure 13A can be generated by equation 7 with various (n, m) combinations. This raises the question of determining whether the double transform variation in Figure 13A is the result of the complex but deterministic process proposed or is the result of random noise.

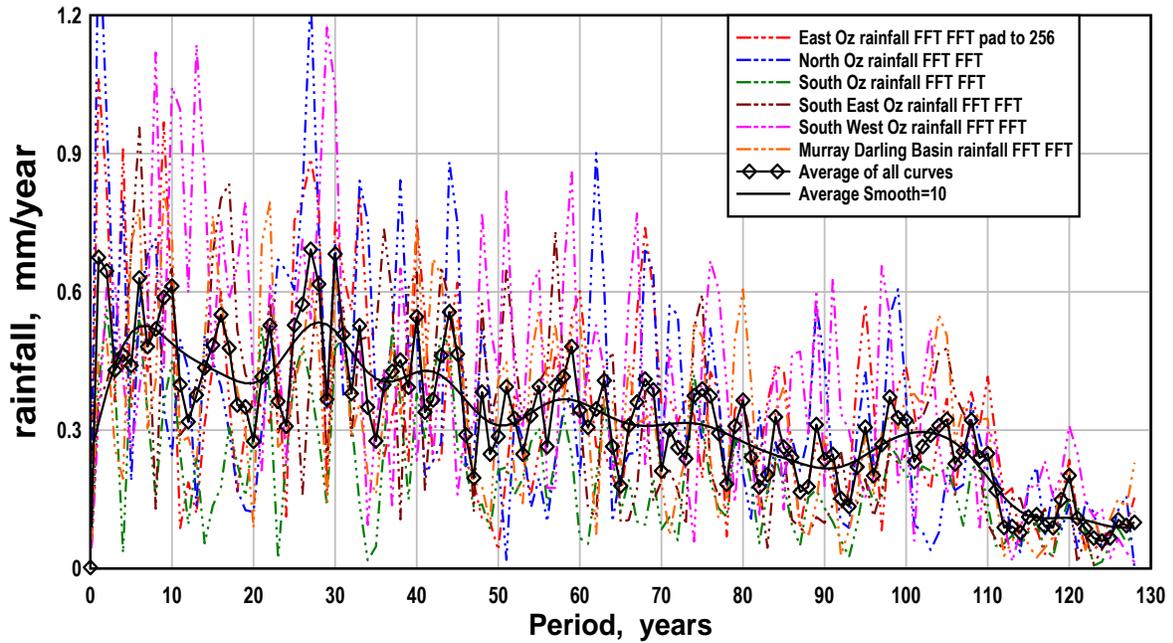

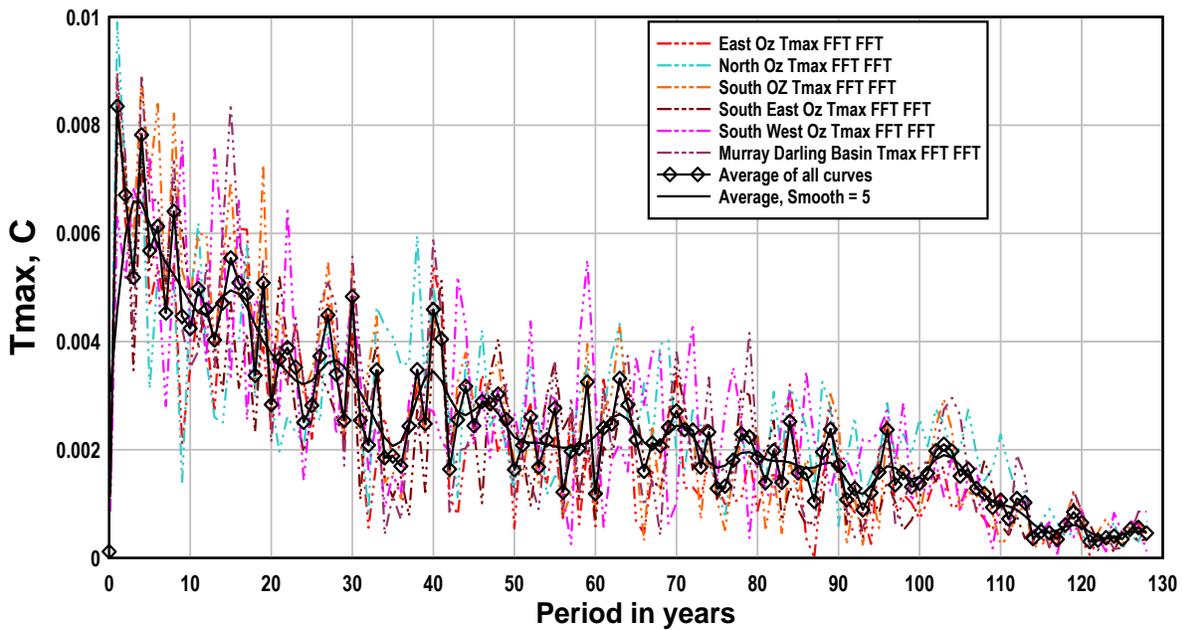

Figure 13. Illustrating the recovery of solar cycle length periodicity from Australian rainfall and temperature data. (A) Shows the second Fourier transforms of Australian regional rainfall data 1900 to 2024. (B) Shows the second Fourier transforms of Australian regional $T_{MAX}$ data 1910 to 2024. The recovered periods are broadly consistent with a dual-tone, (30 year, 100 year), frequency modulated ENSO influencing the variability in Australian rainfall and maximum temperature.

**7.3.2 Solar frequency modulation in the Central England Temperature record.** In the concept proposed here it is expected that most of the different oceanic/atmospheric oscillations on the planet will be frequency modulated due to forcing by solar cycle period variability. Thus the spectral content of a climate variable in any region influenced by a specific oscillation is expected to reflect the predominant component of solar cycle length periodicity that prevails in the same time interval as the climate record. The Australian region is influenced predominantly by ENSO whereas Western Europe is influenced predominantly by the North Atlantic Oscillation, (Olsen et al 2012, Drews et al 2022). The Central England Temperature record extends from 1659 to the present and is available as a daily, monthly or annual data file. Here we use the annual data as organised into season by the Met Office, https://www.metoffice.gov.uk/hadobs/hadcet/data/v2_0_0_0/meantemp_seasonal_totals.txt. Over the 362 year long CET record the period, P, of the solar cycle length will vary considerably. Referring to Figure 7 we notice that the ~ 30 year modulation component is substantially stronger than the other modulation components during the 1659 to 1800 interval. Here we attempt to recover the ~ 30 year solar modulation period from the CET data during this interval using the double Fourier transform method. The first Fourier transform spectra of the CET during the 1659 to 1800 interval and for the four different seasons are shown in Figure 14A. Clearly the greatest variance in temperature is during the December, January, February (DJF) season, as previously reported, (Vertenenko and Dimitriev 2023, Vertenenko and Ogurstov 2012). So we analyse the CET in the DJF season by removing the mean from the CET DJF spectrum and, after performing a second Fourier transform, we recover the modulation period spectrum, Figure 14B. As expected, the power is centred mainly around 30 years with the five year running average peaking at 32 years. Thus the predominant period observed in the solar cycle length variation, 1659 – 1800, Figure 7, is also the predominant frequency modulated component recovered from the CET record. The mean solar cycle period, $T_{M0}$, during 1659 – 1800 is 10.7 years as determined from Figure 6B. This period is shown as the solid reference line in Figure 14A. Also shown are dotted reference lines at the frequencies, $f_{M0}$ +/- $nf_P$, where $f_{M0}$ = 1/10.7 $yr^{-1}$ and $f_P$ = 1/32 $yr^{-1}$. It is evident that the major peaks in the CET DJF first Fourier spectrum coincide fairly closely with the $f_{M0}$ +/- $nf_P$ frequencies. This is most evident towards the higher frequency end of the spectrum as expected by comparison with the modelled spectra, Figure 11B, that shows more consistent spacing of spectral peaks at higher frequencies.

When the same procedure is carried out for the entire 1659 – 2023 CET DJF record a much wider range of frequency modulation periods are recovered. Figure 14C shows that the modulation period 32 years is still prominent. However, other prominent modulation periods, are evident at 15, 77, 165 and 203 years.

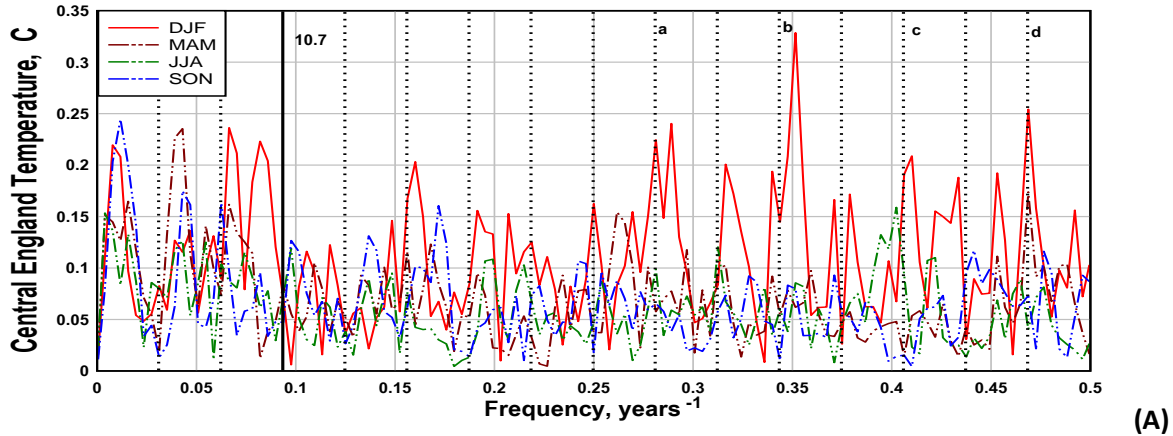

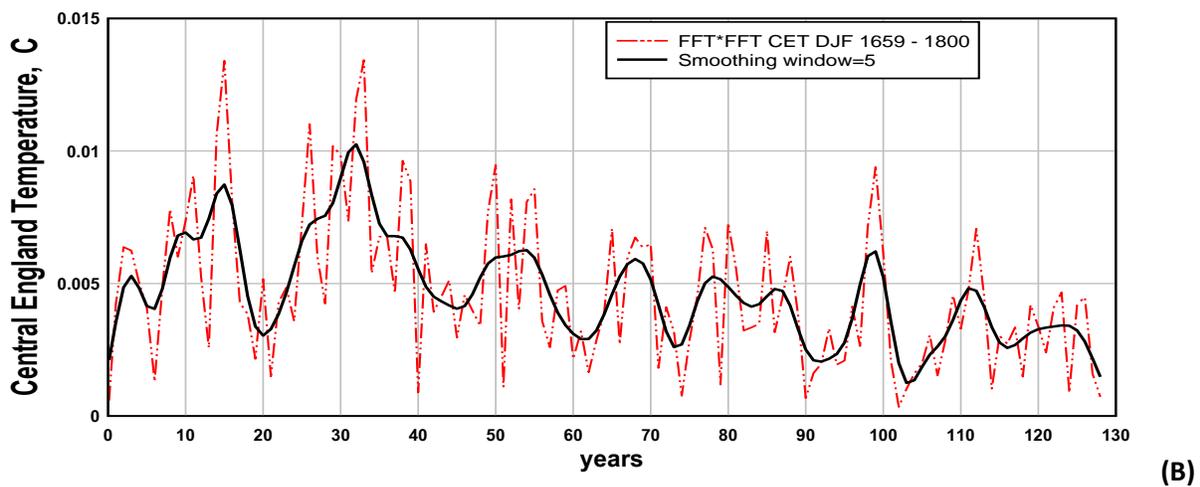

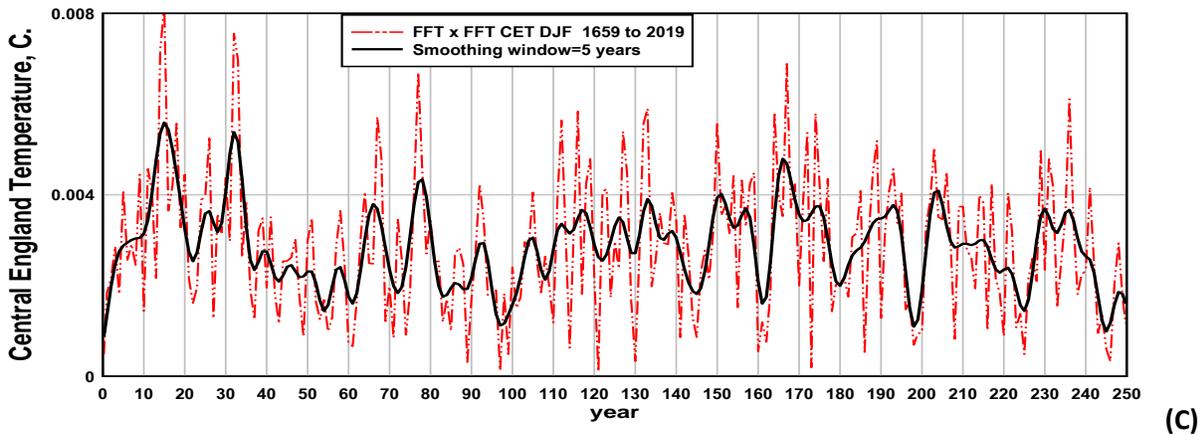

Figure 14. Illustrates recovery of modulation periodicity from CET data. (A) The first Fourier transforms of CET seasonal data 1659 – 1800. The reference lines are spaced at frequencies 1/32 = 0.031 years$^{-1}$. (B) The second Fourier transform of the DJF season CET data recovers a prominent modulation periodicity at ~ 32 years. (C) The double Fourier transform applied to the entire DJF CET record recovers a wide range of modulation periodicity including the periodicity at ~ 32 years.

**7.3.3 Solar cycle frequency modulation in the ENSO SST record.** The result of performing a double Fourier transform on the ENSO SST record 1848 to 2010 is shown in Figure 15A and shows several prominent modulation periods: 20, 30, 49 years and 90 years. Previously, based on an estimate of dominant solar cycle length variation, Figure 7, the dominant solar cycle modulation period, P, was taken to be 100 years. With the double FFT providing an empirical method of finding modulation period a better long period estimate, P = 90 years, is recovered. It is interesting that with the substitution, P = 90 years for P = 100 years, and no other changes to the ENSO model parameters we still obtain a close replication of the observed time variation of the 11 year component in ENSO with the model, Figure 15B

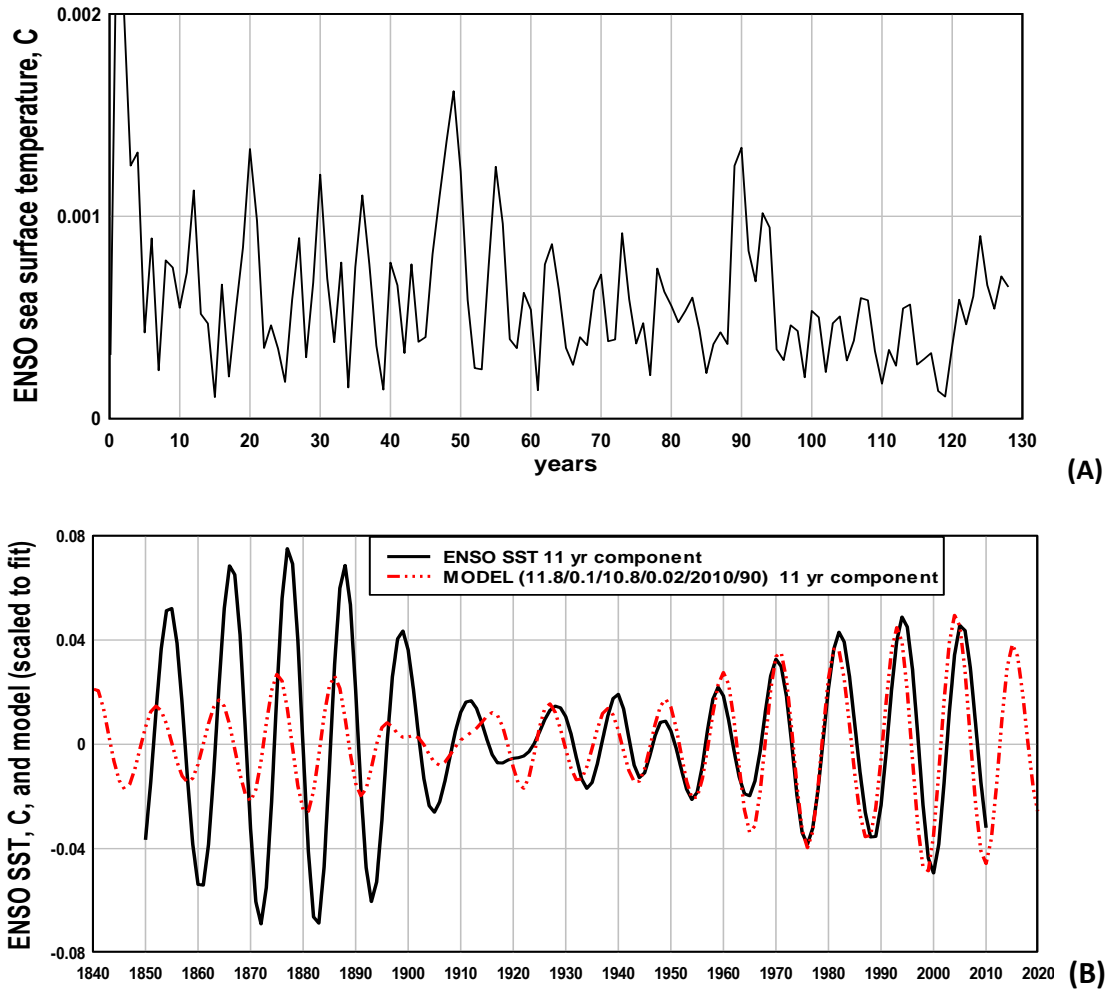

**Figure 15.** (A) The double Fourier transform of the ENSO SST record 1848 – 2010 shows prominent modulation periods at 20, 30, 49 and 90 years. (B) The time variation of the model ENSO variation when the single tone period is taken as P = 90 years and there are no other changes to model parameters.

**7.3.4. Solar cycle modulation periods in reconstructed ENSO variability.** Reconstructions from tree ring growth widths are commonly used to reconstruct historical ENSO variability, for example, Gergis and Fowler (2009). One of the longest reconstructions was obtained by Li et al (2011) who used the North American Drought Atlas, a database of drought reconstructions based on tree-ring records, to produce an annually resolved record of ENSO SST variability over the past 1,100 years, from 900 to 2002, that can be downloaded at https://www.ncei.noaa.gov/pub/data/paleo/treering/reconstructions/enso-li2011.txt .

Here we use that record to recover solar cycle modulation periods from the ENSO reconstruction during intervals that, according to the solar cycle length anomaly of Figure 7, appear to be dominated by one solar cycle length period. We select two intervals, the interval 1400 - 1500 as this interval is expected to be dominated by a ~ 30 year modulation period and the 1020 - 1160 interval as this interval appears to be dominated by long modulation periodicity. The mean modulation period over each of these intervals was estimated visually from Figure 6 to be $T_{M0}$ = 12.0 years for the 1400 – 1500 year interval and $T_{M0}$ = 12.5 years for the 1020 - 1160 year interval. Figures 16A to 16D illustrate the results of the solar cycle modulation period recovery using the double Fourier transform method. We Fourier transform the data in each time interval to obtain the first Fourier transforms, Figures 16A and 16C. Note that the low frequencies in the first Fourier spectrums are attenuated because Li et al (2011) applied a nine year high pass filter to the reconstructed ENSO values. The mean values of the first Fourier transforms are subtracted from each transform and a second Fourier transform applied to the results giving Figures 16B and 16D. As expected a broad peak at P ~ 30 year period was evident in reconstructed ENSO during the 1400 to 1500 interval, Figure 16B, as well as a peak close to the sub harmonic at 60 years.. During the 1020 to 1160 interval there was a broad peak at P ~ 70 years, Figure 16D. We then return to the first Fourier transforms and place reference lines at $f_{M0}$ +/- $nf_P$, (n = 1, 2, 3, etc), in the respective first Fourier transform graphs, Figures 16A and 16C. It is clear that the spacing of the peaks in the spectrums reflects the predominance of short period modulation, P ~ 30 years, corresponding to wide peak spacing 1/30 year$^{-1}$, during the 1400 – 1500 interval and the predominance of longer period modulation, P ~ 70 years, corresponding to narrow peak spacing 1/70 year$^{-1}$, during the 1020 to 1160 interval.

## 8. Discussion and conclusion

It is well known that climate variables like rainfall are highly correlated with the large scale oceanic – atmospheric oscillations like the ENSO and the NAO. For example, in this work, Figure 5 and Figure 10 show the close correlation, (or close anti-correlation) of the ~11 year period ENSO component with Australian cloud, rainfall and temperature ~11 year components. So, an understanding of the origin of climate variability depends largely on understanding the origins of the variability of large scale oscillations like ENSO and NAO.

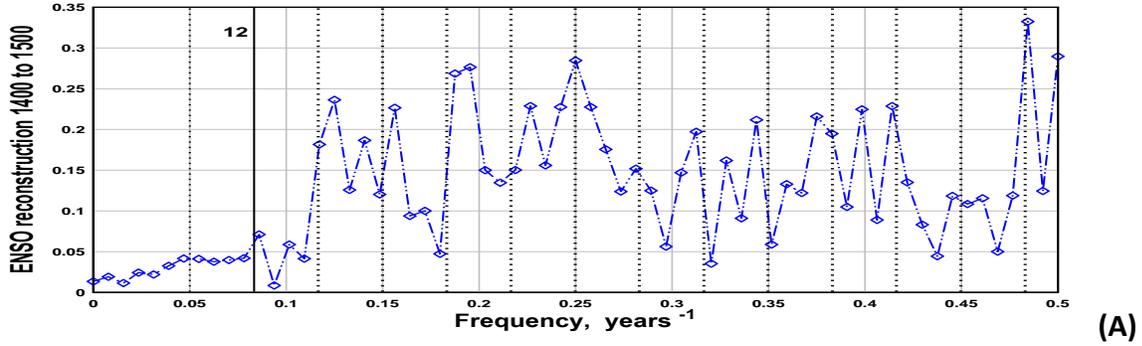

(A)

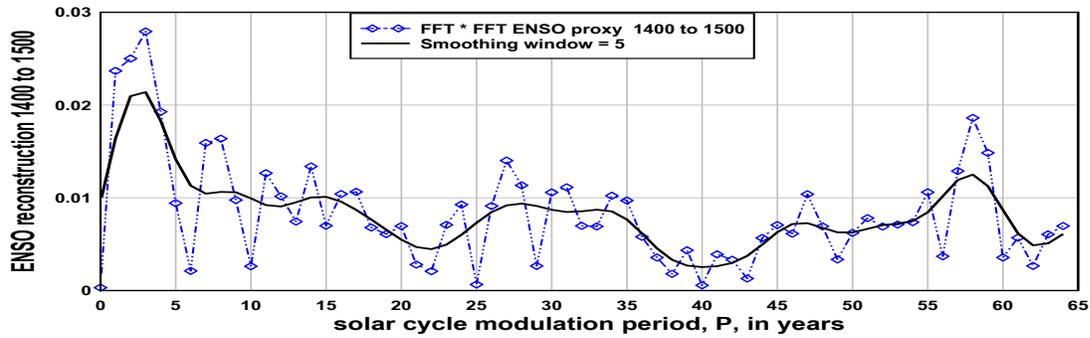

(B)

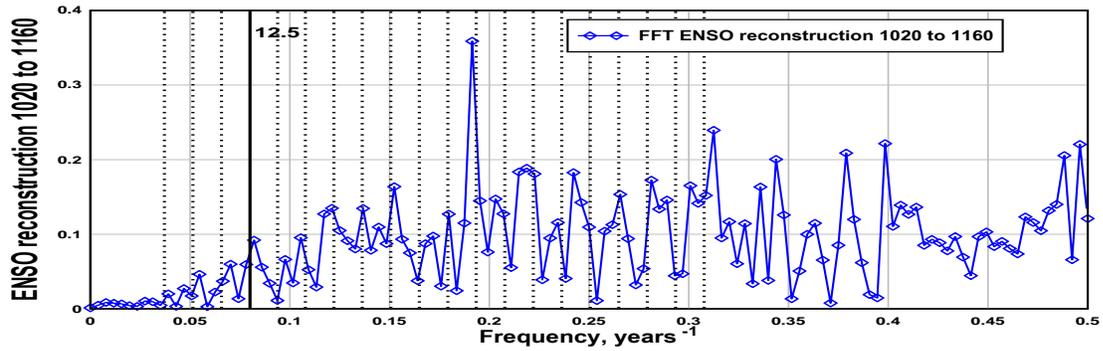

(C)

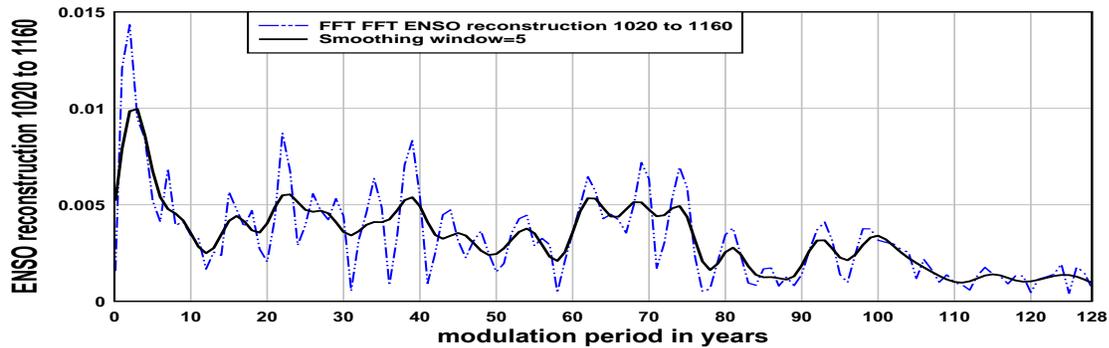

(D)

**Figure 16.** (A) and (C) First Fourier transforms of reconstructed ENSO for intervals 1400 to 1500 and 1020 to 1160 respectively. The reference lines are spaced at Δf = 1/30 = 0.033 yr$^{-1}$ in (A) and at 1/70 = 0.014 yr$^{-1}$ in (C). (B) and (D) Second Fourier transforms of reconstructed ENSO recover a prominent modulation period at ~ 30 years for interval 1400 to 1500 and recover a prominent modulation period of ~ 70 years for interval 1020 to 1160.

The focus in this paper has been on a model of ENSO variability due to frequency modulation of the oscillation by variations in solar cycle length. The history of understanding ENSO variability has developed over several different models or paradigms. The stochastic climate model paradigm, reviewed by Deser et al (2010), holds that ENSO variability is forced by short term, ~ a week long, random atmospheric variability, essentially white noise forcing due to weather events. The resulting short term transfer of heat to or from the sea is integrated in the upper ocean layer, resulting in a longer term red noise-like variability of ENSO sea surface temperature. This stochastic model is consistent with the apparent quasi-periodic variability of ENSO. However, in the spectral content of red noise amplitudes at lower frequencies are much stronger than amplitudes at higher frequencies. This is not consistent with the observed spectra of ENSO and climate variability generally that tends to exhibit flat spectra, i.e. spectra with low and high frequency amplitudes of about equal strength, e.g. Tourre et al (2001). Similarly, in this paper, Figures 3A, 9 and 28 show essentially flat spectra for ENSO, rainfall and temperature variability. Also, evidence, Bruun et al (2017), indicates that ENSO exhibits distinct long term and short term, roughly equal strength, climate cycles with resonant periods of about 2.5, 3.8, 5, 12 – 14, 61 -75, and 180 years. To explain the presence of resonant periods delayed oscillator models of ENSO that were not stochastically forced, (White et al 2001, Bruun et al (2017), were developed. When it became evident that the decadal oscillation in ENSO was coherent with solar activity during the 20th century, models of ENSO as a damped resonant oscillator forced by the amplitude of solar activity were proposed, White and Liu (2008), see also Wasko and Sharma (2009 A & B). However, observations show that ENSO and solar activity are in anti-phase in the 19$^{th}$ century. Thus, unless some process that generates regular and slow phase reversals of ENSO and solar activity can be invoked, it is clear that ENSO is not directly forced by the amplitude of solar activity.

The model of ENSO variability outlined here, frequency modulation of ENSO by the variation in solar cycle frequency avoids the shortcomings of earlier models. With frequency modulation of ENSO the spectrum of ENSO variability is expected to be broad band and flat, as indicated in the modelling results of Figure 8D and Figure 11A, and a flat, broad band spectrum is in accordance with observation. Depending on the dominant period, P, of the solar cycle length variation during the interval of observation the ENSO spectral content will evidence peaks at frequency spacing 1/P. Previously such peaks were construed as multiple resonant periodicities, for example Bruun et al (2017) rather than an intrinsic manifestation of frequency modulation of ENSO. However, the spacing of peaks will vary with time depending on how the dominant periodicity of the solar cycle length varies. For example, Figure 14A shows peaks in the spectra of CET occurring at intervals of 1/32 years$^{-1}$ during the first half of the CET record – corresponding to a solar cycle length period of 32 years being dominant during this time.

The observed ENSO time variation, Figure 4B, is very complex, a combination of many frequency components. In contrast the time variation of the ENSO model appears relatively simple, Figure 8C. In this article we have focused on comparing decadal components, for example Figure 9. While comparison of decadal components is quite convincing a broader comparison is desirable. By using two of the major components of ENSO, the 11 year and 5.7 year components from Figure 4B, a closer approximation of observed ENSO can be compared with the model ENSO time variation, Figure 17.

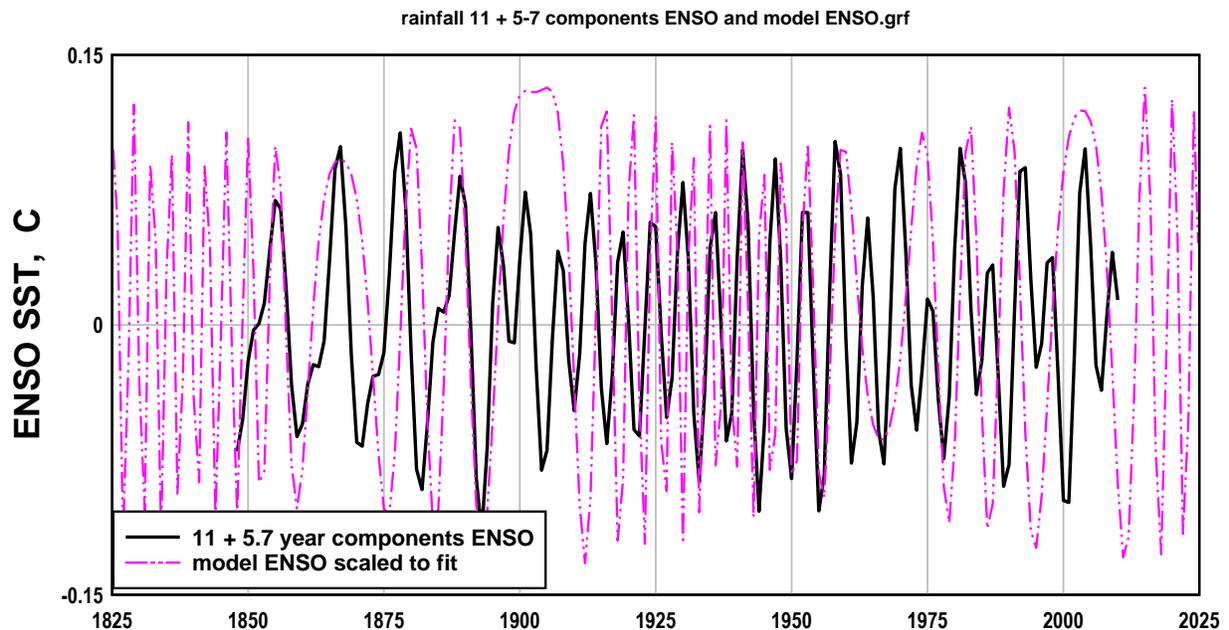

**Figure 17.** The two component version of observed ENSO, black full curve, shows more clearly that the 100 year period frequency modulated ENSO model, pink broken curve, corresponds quite closely to an underlying cycle between strong, long return interval and weaker, short return interval El Nino and La Nina events

In conclusion, a frequency modulated model of ENSO variability advances understanding of the connection between solar variability and climate variability. The model provides several methods for relating solar cycle length variability to climate variability, for example the double Fourier transform method of recovering solar cycle length periodicity from climate variations. However, the actual physical mechanism by which variation in solar cycle length influences ENSO oscillatory response is outside the scope of the paper. Physical mechanisms connecting solar and climate variation have been proposed elsewhere, for example (Gray et al 2010, Vertenenko and Dimitriev 2023). As the interval of most instrumental climate records is short, about 150 years, while solar cycle periodicity is primarily in the 30 to 300 year range, Figure 6C, it follows that future assessment of the model outlined here should focus on using the longest instrumental climate records and on use of the extensive range of paleo-climate records now available, https://www.ncei.noaa.gov/pub/data/paleo/treering .

**Acknowledgements.** The development of the concept outlined here was greatly facilitated by the prior work of Usoskin et al (2021) in reconstructing solar cycle lengths from tree ring data, by Allen et al (2020) in reconstructing precipitation data, the BOM in providing Australian rainfall, cloud and temperature data and the Met Office for providing CET data. I acknowledge useful discussions with Dr Peter Killen and Dr Glen Johnston on the development of the inverted notch filter method.

**References.**